\documentclass[12pt, draftclsnofoot, onecolumn]{IEEEtran}
\usepackage{cite}

\def\bb0{{\mathbb{0}}}

\def\bb{{\mathbf{b}}}

\def\b0{{\mathbf{0}}}

\def\bF{{\mathbf{F}}}

\def\sf0{{\mathsf{0}}}

\def\kron{\otimes}

\usepackage{graphicx}
\usepackage{array}
\usepackage{tikz}
\usepackage{tikz-3dplot}
\usetikzlibrary{shapes.geometric}
\usetikzlibrary{decorations.pathreplacing}
\usetikzlibrary{positioning}
\usepackage{pgfplots}
\usepackage{pstool}  
\usepackage{tikz}
\usetikzlibrary{automata,positioning}
\usepackage[linesnumbered,algoruled,boxed,lined]{algorithm2e}
\usepackage[absolute]{textpos}
\usepackage{threeparttable}
\usepackage[strict]{changepage}
\usepackage{bm,upgreek}
\usepackage{amssymb}
\usepackage{amsmath}
\usepackage{comment}
\usepackage{bbm}
\usepackage{hhline}
\usepackage{amsthm}

\usepackage{lipsum} 
\usepackage{enumitem}
\usepackage{soul}
\usepackage{epsfig}
\usepackage{epstopdf}

\usepackage{amsmath}
\ifCLASSOPTIONcompsoc
  \usepackage[caption=false,font=normalsize,labelfont=sf,textfont=sf]{subfig}
\else
  \usepackage[caption=false,font=footnotesize]{subfig}
\fi

\pgfplotsset{compat=1.14}

\newcommand{\review}[1]{\textcolor{black}{#1}}
\newcommand{\Rev}[1]{\textcolor{black}{#1}}
\begin{document}

\title{Deep Learning Predictive Band Switching in Wireless Networks}

\author{Faris~B.~Mismar, 
        Ahmad~AlAmmouri, 
        Ahmed~Alkhateeb, 
        Jeffrey~G.~Andrews, 
        and~Brian~L.~Evans
\thanks{F.\ B.\ Mismar, A.\ AlAmmouri, J.\ G.\ Andrews, and B.\ L.\ Evans are with the Wireless Networking and Communications Group, Dept. of Electrical and Comp. Eng., The University of Texas at Austin, Austin, TX, 78712, USA. e-mail: \{faris.mismar, alammouri\}@utexas.edu and \{jandrews, bevans\}@ece.utexas.edu.  A.\ Alkhateeb is with the School of Electrical, Computer and Energy Engineering at Arizona State University, Tempe, AZ 85287, USA. email: alkhateeb@asu.edu.}%
\thanks{A preliminary version of this work was presented at the 2018 International Conference on Communications Workshops \cite{Partially_Mismar18}.}%
}

\maketitle
\begin{abstract}
In cellular systems, the user equipment (UE) can request a change in the frequency band when its rate drops below a threshold on the current band.  The UE is then instructed by the base station (BS) to measure the quality of candidate bands, which requires a \emph{measurement gap} in the data transmission, thus lowering the data rate.  We propose {an online-learning based} band switching approach that does not require any measurement gap.  Our proposed classifier-based band switching policy instead exploits spatial and spectral correlation between radio frequency signals in different bands based on knowledge of the UE location.   We focus on switching between a lower (e.g., 3.5 GHz) band and a millimeter wave band (e.g., 28 GHz), and design and evaluate two classification models that are trained on a ray-tracing dataset.  A key insight is that measurement gaps are overkill, in that only the relative order of the bands is necessary for band selection, rather than a full channel estimate.  Our proposed machine learning-based policies achieve roughly 30\% improvement in mean effective rates over those of the industry standard policy, while achieving misclassification errors well below 0.5\% {and maintaining resilience against blockage uncertainty}.
\end{abstract}

\IEEEpeerreviewmaketitle

\section{Introduction}
With each successive cellular standard using a rapidly increasing number of different frequency bands in different parts of the spectrum, the band selection problem has become ever more complicated.   In particular, user equipment (UEs) would like to use the band or bands that maximize their quality of experience (QoE), which is highly correlated to their achieved data rate.   The choice of the optimal frequency band can be challenging.  On the one hand, lower frequency bands generally have more benign propagation properties and thus produce higher signal to noise ratios (SNRs), but higher frequency bands such as millimeter wave (mmWave) offer much higher bandwidth as well as beamforming gains and will typically be more lightly loaded.  So, if the SNR on a mmWave band is acceptable, it is likely to provide a much higher data rate than a lower band and a UE would usually benefit from being efficiently switched over to the mmWave band. Similarly, if coverage is lost on the mmWave band, the UE should be quickly switched back to the lower frequency band.

\begin{figure}[!t]
\centering
\begin{tikzpicture}[node distance=3cm] 


\node[draw, rectangle] (UE)  at (-2,0) {UE};
\node[draw, rectangle, right of=UE,xshift=9cm] (BS1) {BS};
\draw (-2,-0.25) -- (-2,-7);
\draw (10,-0.25) -- (10,-7);

\draw[-, >=latex, ultra thick] (-2,-1.5)  -- node [text width=10cm, align=center, above] {Received power of the user \review{on} $f_j$ drops below threshold} ++(12,0);
\draw[->, >=latex, ultra thick] (-2,-2.5)  -- node [midway,above]{Request band switch to $f_{j^\prime}$} ++(12,0);
\draw[<-, black, >=latex, ultra thick] (-2,-3.5)  -- node [text width=12cm, align=center, above] {Measurement gap configured at $f_j$}  ++(12,0);
 
 \draw[-, black, >=latex, ultra thick] (-2,-4.5)  -- node [text width=6cm, align=center, midway,above]  {Measure the new channel at $f_{j^\prime}$} ++(12,0);
\draw[->, >=latex, ultra thick] (-2,-5.5)  -- node [text width=12cm, align=center, midway,above] {Report the measurements} ++(12,0);
 \draw[<-, >=latex, ultra thick] (-2,-6.5)  -- node [text width=12cm, align=center, midway,above]{Band switch decision}  ++(12,0);
\end{tikzpicture}
\caption{The band switch procedure  between frequency bands in one base station (BS) \review{in the downlink direction}.}
\label{fig:ho_proc}
\end{figure}
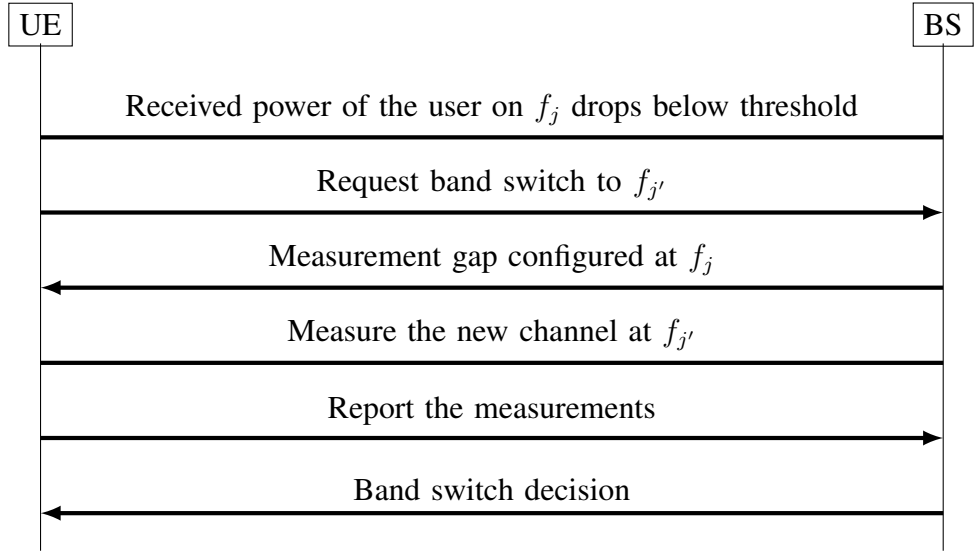

Despite its increasing importance, the signaling procedure for band switching has seen only incremental changes over the \review{evolution of}  multiple successive 3GPP standards  \cite{3gpp36331,3GPP2007R8}. This signaling {(or \textit{control plane})} procedure is shown in Fig.~\ref{fig:ho_proc} and described as follows. If the received power at the UE drops below a certain threshold on its current frequency band, call it $f_j$, it requests a band switch from its serving base station (BS). This request is followed by a \emph{measurement gap}, where the data {(or \textit{user plane})} flow is stopped to allow the user to tune its reception circuitry to the frequency of the target band, call it $f_{j^\prime}, j^\prime\neq j$, to measure the channel. {The industry standards justify ceasing the data flow to preserve the UE battery.} After obtaining the measurements, the user reports them back to the BS.  The BS estimates, based on the measurements, whether the user would benefit from switching to $f_{j^\prime}$ or not, and hence, grants or denies the request. \review{A key} issue with the aforementioned procedure is its dependence on the measurement gap which {despite being a control plane procedure} causes interruption in the flow in the {data plane} and reduces the user overall throughput. \review{The 3GPP standards also introduced mobile load balancing (MLB) as a means of transferring traffic served by a congested BS to nearby BSs that have spare resources \cite{3gpp32860}.  However, MLB requires periodic communication between BSs about their resources, introducing a significant overhead.  MLB is also triggered by the BS desire to relieve its congestion, while the band switch is triggered when the UE desires to maximize its QoE.}

\review{It would be desirable to introduce a reliable} method that can support the band switch procedure without interrupting the user data flow by measurement gaps. The aim of this paper is to propose \review{a novel} {online-learning based} gap-free algorithm for band switching that utilizes the spatial and spectral correlations over different frequency bands along with the previous band switching requests and decisions for nearby users.  {This online-learning based algorithm adapts to the changes in the environment as experienced by the users.}  Precisely, we propose a predictive algorithm, based on deep neural network (DNN) classifiers, which allow the BS to decide whether to grant or deny the band switching requests without the need for measurement gaps.

\subsection{Related Work}
Predicting the success of a band switch from one frequency band to another without explicitly measuring the channel at the target frequency band falls under the genre of problems commonly referred to as: ``channel estimation using out-of-band information'' \cite{Millimeter_Gonzalez17}. In the simplest form of this problem, there are forward and backward (downlink and uplink) links occupying the same frequency bands at different time slots. In this case, we can use channel reciprocity \cite{Spatial_Hugl02} to estimate the channel of the backward link using the measurements on the forward link, or vice versa. Even with a separation of frequency duplex bands on the order  of ten megahertz, a spatial correlation between the signals on the two frequency bands still exists due to the common propagation paths, blockages, and reflectors \cite{Spatial_Hugl02, Deep_Alrabeiah19}. Interestingly, the spatial correlation between two frequency bands that are separated by tens of gigahertz still exits \cite{Nitsche_Steering15}. However, it cannot be directly used to accurately estimate the channel on one frequency band by only using the measurements from the other, but it can be used to aid the channel estimation and reduce its complexity. For example, this correlation was exploited in \cite{Millimeter_Ali18,Xiu_A18,Xiu_Millimeter19,Spatial_Ali19} for cell discovery, channel covariance estimation, and beam selection in mmWave bands using sub-6 GHz measurements. In the case of band switching, we are not interested in accurate channel estimation since the objective is not to use the estimate in decoding the messages, beam selection, or precoder/combiner design used in multiple-input multiple-output (MIMO) communications. Instead, our \review{goal is much simpler: \textit{ranking}} the downlink channel quality of the two frequency bands or technologies.

The major challenge in exploiting the spatial correlation between frequency bands is the lack of accurate mathematical models that describe how the channel changes across these frequencies (or technologies). This challenge makes a data-driven or a machine learning (ML) approach more attractive to follow and implement.   %
With more publicly available datasets that are based on field-measurements or sophisticated ray-tracing simulations \cite{DeepMIMO_Alkhateeb19, deepmimo_code}, we expect the interest in this approach to dramatically increase. \review{Nevertheless,} the applications involving dual-band ray-tracing datasets with ML classification to study channels is a nascent research area. 

Although relevant to dual-band resource management, the work in \cite{Joint_Semiari17}  did not address the impact of measurement gaps on UE data rates.  It focused on granting resources to users at mmWave first, while we allow granting resources to both mmWave and sub-6 GHz simultaneously without any specific preference.  Furthermore, statistical path loss models were used for both mmWave and sub-6 GHz bands which may be privy to the spatial and spectral correlation of channels that we otherwise capture using ray-tracing datasets.  

The work in \cite{Caching_Semiari18} studied only one type of 3GPP dual-band handovers, which we call the ``legacy'' policy later in this paper.   However, similar to \cite{Joint_Semiari17}, the use of statistical path loss models voids the opportunity to exploit the correlation across bands; therefore insights about the performance of the various algorithms, including the second type of 3GPP dual-band handover algorithms---the ``blind'' policy---could not be derived.  Furthermore, the objective was to improve energy efficiency through handover avoidance, unlike our proposed algorithm the objective of which is to improve the UE data rates by eliminating measurement gaps.

In \cite{Improved_Polese17} dual connectivity was studied.   Dual connectivity requires a local coordinator to manage the traffic between the cells, unlike band switch procedures.  As a result, a backhaul latency constraint between the BSs was imposed.  Furthermore, empirical pathloss models were used. Multiple BSs with a single UE were simulated while our focus on a single BS with dual band and multiple UEs.  The use of a single UE may prevent the employment of ML techniques due to the limited number of learning observations---a problem we avoid altogether through the use of a ray-tracing dataset of RF propagation paths from UEs to a BS.  Moreover, a band switch time-to-trigger mechanism was introduced in \cite{Improved_Polese17}, whereby the band switch is only granted after the band switch criterion is fulfilled for a period of time.  This, unlike our proposed approach, introduces further latency to the band switch procedure \cite{What_Andrews14}.

\subsection{Contributions}
\review{In this paper, we provide an answer to the question whether a reliable band switch method exists to maximize the users' achievable data rates.  Specifically, this paper makes the following contributions:}

\begin{enumerate}
    \item Motivate the use of deep learning in ranking the downlink channel quality of the two frequency bands---a mathematically intractable problem and a requirement for the band switching procedure.
    \item Offer several insights about the different band switch policies and their respective impact on performance.   Furthermore, we show how the choice of the band switch threshold can have adverse impacts on the performance.
    \item Motivate a data-driven approach to band switching, where we use a ray-tracing dataset in deep learning.
    \item Create a unified framework to describe the band switch policies in a single equation and use this equation to explain the various band switching policies and their relevant performance. 
\end{enumerate}%
\section{System Model}\label{sec:system}

In this section, we describe the adopted network and channel models. 

\subsection{Network Model}
We consider a radio network comprised of one BS serving single-antenna user equipment (UEs) in an arbitrary association area. The BS has two frequency bands; one in the sub-6 range and one at mmWave. {Note that we assume that sub-6 and mmWave BSs are co-located to minimize the financial costs of deployment}.  Moreover, the BS utilizes a different number of antennas for each frequency band. Let $j\in \{\text{sub-6}, \text{mmWave}\}$ denote the frequency band and let $N^{(j)}$ denote the number of antennas on the $j^\text{th}$ band, then the received signal at the $i^\text{th}$ UE from the BS at the $j^\text{th}$ frequency band is
\begin{align}\label{eq:channel}
    r_{(i,j)} = P_\text{TX}^{(j)}\mathbf{h}_{(i,j)}^\ast\mathbf{f}_{(i,j)} s_{(i,j)} + n_{(i,j)},
\end{align}
where $P_\text{TX}^{(j)}$ is the transmit power of BS on the $j^\text{th}$ frequency band, $\mathbf{h}_{(i,j)}\in\mathbb{C}^{N^{(j)}\times 1}$ is the channel vector, $\mathbf{f}_{(i,j)}$ is the  beamforming vector,  $s_{(i,j)}$ is the transmitted signal, and $n_{(i,j)}\sim \mathrm{Normal}(0, \sigma_j^2)$ is the thermal noise {computed over the bandwidth $B^{(j)}$} {including the UE noise figure}. We focus on codebook-based analog beamforming, where the beamforming vector is chosen from a pre-defined codebook $\mathcal{F}^{(j)}$ \cite{Beam_Jianhua19}. In this case, the BS chooses the optimal beamforming vector $\mathbf{f}^\star$ that maximizes the receive SNR from the $i^{\rm th}$ user on the $j^{\rm th}$ frequency band from the codebook $\mathcal{F}^{(j)}$
\begin{equation}
    \mathbf{f}_{(i,j)}^\star := \underset{\mathbf{f}_{(i,j)}\in\mathcal{F}^{(j)}}{\arg\, \max} \  \vert\mathbf{h}_{(i,j)}^*\mathbf{f}_{(i,j)}\vert^2.
\end{equation}

Let the codebook size be denoted by $N_\text{CB}^{(j)}$ and assume that all codewords are normalized, i.e., $\Vert \mathbf{f}_{(i,j)}\Vert ^2=1$. Based on this, the received SNR at time step $t$ at the $i^\text{th}$ UE on the $j^\text{th}$ frequency band is
\begin{equation}
    \begin{aligned}
    \gamma^{(i,j)}[t] = \frac{P_\text{TX}^{(j)}[t]}{\sigma_j^2} \vert \mathbf{h}_{(i,j)}^{\ast}[t]\mathbf{f}_{(i,j)}^{\star}[t] \vert^2,
    \end{aligned}
    \label{eq:snr_ifho}
\end{equation}
and the instantaneous achievable rate is
\begin{equation}
R^{(i,j)}[t] = B^{(j)}\log_2 (1+\gamma^{(i,j)}[t]),     
\label{eq:shannon}
\end{equation}
where $B^{(j)}$ is the available bandwidth at the $j^\text{th}$ frequency. Note that the rate in \eqref{eq:shannon} does not include the overhead of switching to a different frequency band nor the beam training overhead. These overheads cause a loss in throughput, which is typically related to the coherence time of the channel and the frame length.

\subsection{Channel Model}
Here we discuss the channel coherence time, the beam training time, the band switching overhead, and the effective throughput.

\textbf{Channel coherence time:} Let the coherence time for sub-6 GHz and mmWave frequency bands be denoted as $T_C^\text{sub-6}$ and $T_C^\text{mmWave}$, respectively.  The exact values depend on the environment, the antenna configuration, and the user movement. Hence, to maintain the generality of the framework, we do not assume specific values for the channel coherence times and we discuss our choices of the coherence times in Section~\ref{sec:dataset}, which is only needed to numerically evaluate the performance of the different algorithms.

\textbf{Beam training time:} For the beam training overhead, we define the training penalty per beam as $T_\text{beam}$.  Thus, the total beam training time, $T_B$, is related to the number of all possible beams, which is the size of the codebook $\mathcal{F}^{(j)}$ in our case (i.e., $T_B=T_\text{beam} N_\text{CB}^{(j)}$).

\textbf{Band switching overhead:} At the beginning of each radio frame, the UE can request a band switch operation from its serving BS if it is not satisfied by its current signal quality.  \Rev{We are using two time slots per frame, and a time slot does not necessarily have an integer duration that is compliant with the standards.  The radio frame duration is not constant and is equal to the channel coherence time.  These band switches can only happen at the beginning in the first time slot with zero delay by design}.  The BS uses a certain policy to determine whether the change to a different frequency band should be granted or denied. However, there is a time penalty for the band switch request, which is used by the BS to take a decision regarding the user request. We denote this overhead by $T_H$, which is determined by the algorithm or the policy used in the BS to respond to the band switch request and the existence or absence of a measurement gap. The exact values of $T_H$ are given in Section~\ref{sec:band_change_policies}, where we present different band switch policies.

\textbf{Effective throughput:} Using the previous definitions for the channel coherence time $T_C$, the beam training time $T_B$, and the band switch overhead $T_H$, we can {define the instantaneous weight $w^{(i,j,k)}[t]$ for the $i^\text{th}$ UE that is connected to the BS on the $j^\text{th}$ frequency band at the time step $t$ as}%

\begin{equation}
    w^{(i,j,k)}[t] := \max \left (0, 1 - \frac{T_B^{(j)}+T_H^{(k)}}{T_C^{(j)}} \right )
\end{equation}
{which accounts for the prolonged band switching overhead as a result of longer measurement gaps, since the throughput cannot be below zero.  Here,} the $j^\text{th}$ band is the band \textit{after} the band switch decision is made, which is a new frequency band if the band switch was granted and the old frequency band if the band switch was denied.  This enables us to compute the effective throughput for {the said UE} as
\begin{align}
      R_E^{(i,j, k)}[t] &= w^{(i,j,k)}[t] R^{(i,j)}[t].
      \label{eq:ear2}
 \end{align}
  
After discussing the system model and providing the necessary definitions, we present the current polices discussed in the industry standards \cite{3gpp36331} for the BS to make band switch decisions in the next section.

\section{Band Switch Policies}\label{sec:band_change_policies}

A band switch policy has to answer the following two questions: (\textit{i}) when should the UE request this band switch? (\textit{ii}) what is the information needed by the BS to make a decision for the band switch request and how? The first is typically solved by a pre-defined rate threshold $r_\text{threshold}$, such that if the UE rate is below this threshold, it requests a band switch. For the second question, the standards specify two policies today \cite{3gpp36331}.  These policies are the measurement-based legacy approach and the blind approach. We also discuss the optimal policy as a benchmark. {This policy is optimal in that it does not require the UE to use a measurement gap, and therefore the UE throughput is at the highest possible given the channel conditions, as we show later in the section.} To provide a unified framework for the different polices, we define the following decision variables: $x_\text{br},{y} \in \{0,1\}$, where $x_\text{br}=1$ if the UE requests a band switch, and $x_\text{br}=0$ otherwise, and $y=1$ if the BS grants the band switch and ${y}=0$ otherwise. It is understood that ${y}$ is only defined if $x_\text{br}=1$.  \review{Further, the threshold  $r_\text{threshold}$ is defined for all policies except the optimal policy.  It is set based on how soon the UE should request the band switch from the BS.}

\begin{figure}[!t]
\centering
\begin{tikzpicture}[node distance=4.5cm, auto, scale=0.9]
\tikzstyle{every node}=[font=\small]

\node[text width=2cm] at (0.6,0.2) {Requests};
\draw[line width=1.5, ->, >=latex] (-0.5,0) -- (8,0) node[below] {time $t$};

\foreach \x in {2.25,3.5,5,6.4}
    \draw[->, >=latex] (\x,0) -- ++(0,0.5);

\foreach \x/\y in {2.25/grant,3.5/deny,5/$\ldots$,6.4/grant} {
    \draw[->, >=latex] (\x+0.5,-1) -- ++(0,0.5) node[above, yshift=-2] {\smaller \y};
    \draw[fill=gray!30] (\x,-1) rectangle ++(0.5,0.25);
};

\node[text width=2cm] at (0.6,-0.8) {Legacy};
\draw[line width=1.5, ->, >=latex] (-0.5,-1) -- (8,-1) node[below] {};

\foreach \x in {2.25,3.5,5,6.4} {
    \draw[->, >=latex] (\x+0.2,-2) -- ++(0,0.5) node[above, yshift=-2] {\smaller grant};
};

\node[text width=2cm] at (0.6,-1.8) {Blind};
\draw[line width=1.5, ->, >=latex] (-0.5,-2) -- (8,-2) node[below] {};

\end{tikzpicture}
\vspace*{-1em}\caption{Legacy band switch timing diagram.  The shaded gray rectangles represent the measurement gaps.}
\label{fig:legacy_diagram}
\end{figure}
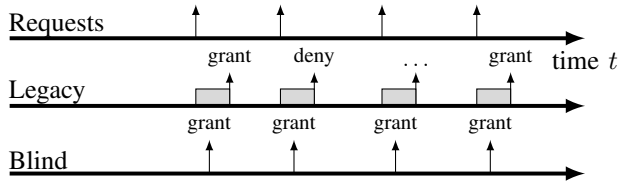

\subsection{Legacy Policy}
The legacy policy, also known as the measurement-based policy, is shown in Fig.~\ref{fig:legacy_diagram}.  When the user throughput is \review{below} the threshold $r_\text{threshold}$, it requests a band switch from the BS and it stops its transmission to measure the channel at the desired frequency band.  After measuring the \review{downlink} channel, the user reports the measurements back to the BS, which decides whether to grant or deny the band switch request based on the measurements provided by the user. The measurement gap duration, denoted by $T_G$, is set to be a fraction of the coherence time\cite{3gpp36331}
\begin{equation}
    T_G := \rho T_C, \qquad 0 < \rho \le 1.
    \label{eq:coherence}
\end{equation}

Further, \review{if we denote the overhead due to a band switch signaling request and its decision response as $\beta > 0$, then the band switch time overhead $T_H^{(\text{legacy})}$ is equal to $T_G + \beta$}.  By using this policy, the BS can make an informed decision regarding the band switch using the rates from both bands, which guarantees a certain QoE for the user. However, this comes at the expense of the measurement gap, where the BS stops its transmission to the UE so it can measure the target channel, which causes an interruption in the data flow and reduces the UE throughput. 

By employing this policy, one of three \review{scenarios are possible} at the beginning of each frame: (\textit{i}) the UE does not request a band switch, which happens if its current rate is higher than the threshold;  (\textit{ii}) the UE requests a band switch and it is granted by the BS, which happens if the user's current rate is lower than the threshold, and the rate at the target band is higher than its current rate; or (\textit{iii}) the UE requests a band switch, but it is denied, which happens if the UE current rate is lower than the threshold, and the rate at the target band is lower than its current rate. For the legacy policy, the decision variables are defined as
\begin{align}
      x_\text{br}^{(i)}[t]&= \mathbbm{1}[{(R^{(i,j)}[t] < r_\text{threshold})}], \qquad \forall i,\label{Eq:LegVar1}\\
       {y}^{(i)}[t]&= \mathbbm{1}[{(\hat{R}^{(i,j^\prime)}[t] > R^{(i,j)}[t] )}], \qquad \forall i\label{Eq:LegVar2},
\end{align}
where $j$ is the current serving BS, $j^\prime$ is the target BS, and $\hat{R}^{(i,j^\prime)}$ is the estimated rate the UE would get if the band switch were granted. 

\subsection{Blind Policy}
Similar to the legacy policy, when the UE throughput is \review{below} the threshold, it requests a band switch from the BS. However, in this policy, the BS instructs the UE to band switch to a different band without any need for a measurement gap.  Given the nature of this band switch approach, if the SNR is worse at the target frequency, the throughput drops significantly. Hence, although the measurement gap is eliminated, the BS cannot guarantee the user a higher throughput after the band switch, which causes a low QoE for the user. The decision variables in this case are as follows
\begin{align}
      x_\text{br}^{(i)}[t]&= \mathbbm{1}[{(R^{(i,j)}[t] < r_\text{threshold})}], \qquad \forall i,\\
       {y}^{(i)}[t]&= 1, \qquad \forall i,
\end{align}
since the band switch requests are always granted by the BS. Here, $T_H^{(\text{blind})} = \beta$, which is only for the signaling overhead since there is no measurement gap requirement.

\subsection{Optimal}
To define an upper bound for the various band switch policies, we define the optimal policy to be the one where the BS knows the instantaneous quality of the channels of the different bands perfectly, so there is no need for a measurement gap. Hence, it asks the user to switch to a different band if the target rate is higher than its current rate. It also eliminates the need for a pre-defined rate threshold, since the band switch request and decision are combined and executed at the beginning of each frame by the BS.   Based on this, the optimal effective throughput in this case is given by
\begin{equation}
    R_E^{\star (i)}[t]  = \underset{\substack{j \in\{\text{sub-6},\text{mmWave}\} }}{\max\;} \left ( 1 - \frac{T_B^{(j)}}{T_C^{(j)}}\right ) R^{(i,j)}[t].
\end{equation}%

Finally, the decision variables can be written as
\begin{align}
      x_\text{br}^{(i)}[t]&= 1 , \qquad \forall i,\\
       {y}^{(i)}[t]&= \mathbbm{1}[{(\hat{R}^{(i,j^\prime)}[t] > R^{(i,j)}[t] )}], \qquad \forall i.
\end{align}

\subsection{Overhead of Band Switching}
Based on the previous discussion, besides the standards-imposed signaling overhead requirement of $\beta$, which is common across all policies, only the legacy policy causes band switch overhead. This overhead is equal to the measurement gap. Hence, $T_H^{(i,j,k)}=T_G^{(i,j,k)} + \beta$ for $k\in\{\text{legacy}\}$ and $T_H^{(i,j,k)}=\beta, k\in\{\text{blind}, \text{optimal}\}$.  Note that deterioration in user throughput due to band switch overhead in the legacy policy drives the setting of the pre-defined threshold to lower values to avoid spending long times in measurement gaps. When this threshold is set low, the signal quality has to be bad for the user to request the band switch.  This prevents the user from utilizing possibly better channels on other frequency bands or technologies. Moreover, with the introduction of mmWave frequency bands \cite{Millimeter_Rappaport13} in the {fifth generation of wireless communications (5G)} standard, the design of the band switch procedure becomes yet more critical since radio frequency signals at mmWave bands are more sensitive to blockages by various objects.  For example, it was shown in  \cite{Hand_Alammouri19} that the antenna gains on the mmWave bands can suffer from up to 25 dB attenuation due to the user hand grip on the mobile device and it varies with the different hand grips.  Hence, under large blockage losses, the user would benefit from a fast transition to other frequency bands, and relying on the measurement gaps does not help. 

The objective of this work is to propose a new band switch policy that eliminates the measurement gap, as in the blind policy, but ensures a certain QoE as in the legacy policy. Our policy relies on deep learning classification, which we introduce in the next section, before discussing the details of our algorithm.

\begin{table*}[!t]
\centering
\setlength\doublerulesep{0.5pt}
\caption{Deep neural network classifier learning features}
\vspace*{-1em}
\label{tab:features}
\begin{tabular}{ l|lll } 
\hhline{====}
& Parameter & Type & Description \\
 \hline
$\mathbf{x_0}$ & Bias term & Integer & This is equal to unity. \\
$\mathbf{x_1}$ & $R_E^{(\text{sub-6})}$ & Float & Effective achievable rate in the sub-6 GHz band.\\
$\mathbf{x_2}$ & $R_E^{(\text{mmWave})}$ & Float & Effective achievable rate in the mmWave band.\\
$\mathbf{x_3}$ & Source technology & Binary & $(= 1$ for sub-6 and $=0$ for mmWave).\\
$(\mathbf{x_4}, \mathbf{x_5}, \mathbf{x_6})$ & Coordinates & Float & The coordinates of the UEs distance from the base station.\\
& & & based on the coordinates of $i^\text{th}$ UE (i.e., $d_i[t]$). \\
$\mathbf{x_7}$ & Band switch requested & Binary & UE requested band switch ($x_\text{br}^{(i)} = 1$)? \\
$\mathbf{y}$ & Band switch decision &  Binary & UE band switch request is granted ($y^{(i)} = 1$)? \\
\hhline{====}
\end{tabular}
\end{table*}


\section{Proposed Policy}\label{sec:proposed}
{For our proposed policy, we define $\mathcal{N}$ as the set of users within the serving BS area.  This set has a cardinality  $\vert\mathcal{N}\vert := T_\text{simulation}N_\text{UE}$.  Here, $T_\text{simulation}$ is the total simulation time and is equal to the number of stratified samples the total set $\mathcal{N}$ is divided into (i.e., partitions).  Also, $N_\text{UE}$ is the number of active UEs within the BS serving area.  Further, we define a set of users $\mathcal{U}$ (also within the serving BS area) with an objective to improve the band switch performance using the proposed policy, but without a measurement gap.}

{To achieve this objective, we} use the locations and measurements of the set $\mathcal{N}\setminus \mathcal{U}$, also served by this base station, such that $\mathcal{U}\subset\mathcal{N}$. This is achieved by exploiting the spatial and spectral correlation of the channels over different frequency bands and different locations.

We keep the minimum threshold criteria used in the legacy and blind polices, where the UE requests a band switch if its rate drops below a pre-defined threshold, $r_\text{threshold}$. Then the BS grants the band switch if the estimated rate at the target frequency is higher than the UE current rate. The difference is that the BS does not ask the UE to interrupt its transmission to measure the channel, but \review{instead} uses a machine learning approach to estimate the rate at the target frequency. Hence, the decision variables $x_\text{br},{y}$ are defined in the same way as in the legacy approach given in \eqref{Eq:LegVar1} and \eqref{Eq:LegVar2}.

 The major challenge in this approach, which relies on exploiting the spatial and spectral correlation between the channels, is the lack of accurate mathematical models that capture these correlations. Hence, we propose the use of DNN classifiers in the solution of our problem.  
 
\begin{figure}[!t]
\centering
\linespread{1.0} 
\begin{tikzpicture}[style=thick,scale=1]
  	\node [rectangle, draw, rounded corners, 
		text width=12em, text centered, minimum height=1em, fill=gray!30, inner sep=10pt, yshift=1.5em] (ho) {Measurement gap free band switch policy};

\path [draw, latex-] (ho.167) -- ++(-1,0) node[text width=12em, align=right, left] {User location};
\path [draw, latex-] (ho.180) -- ++(-1,0) node[text width=12em, align=right, left] {Current frequency band};
\path [draw, latex-] (ho.193) -- ++(-1,0) node[text width=12em, align=right, left] {Current achievable rate};

	\path [draw, latex-] (ho.210) -- ++(0,-1) node[text width=8em, align=center, below] {Band switch\\ requests};
	
	\path [draw, latex-] (ho.330) -- ++(0,-1) node[text width=8em, align=center, below] {Prior decisions};

	\path [draw, -latex] (ho.0) -- ++(1,0) node[text width=12em, align=left, right] (a) {Band switch decision};
\end{tikzpicture}%
\vspace*{-1em}\caption{Illustration of our proposed algorithm.  The list of learning features is shown in Table~\ref{tab:features}.}\label{fig:ifho_overall}
\end{figure}
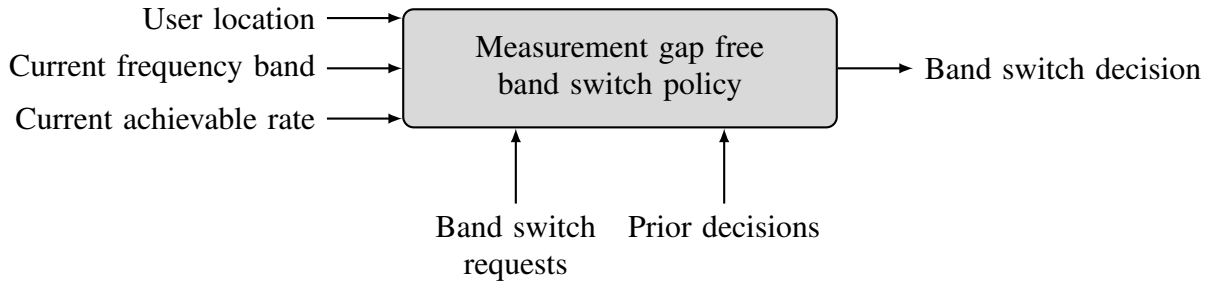

 We use the classifiers to predict the band switch of other locations in the proximity of these learned locations.  
 Our algorithm is illustrated in Fig.~\ref{fig:ifho_overall} and is specified in Algorithm~\ref{alg:algorithm}. The main steps of Algorithm~\ref{alg:algorithm} are as follows:
\begin{itemize}
    \item Construct the dataset for the UEs {within the serving BS area} {in the current time step $t$}, which contains the rates from the spatially correlated wireless channels and band switch decisions {and the stratified sets\footnote{{Given that the set $\mathcal{N}$ is stratified, it should be clear that $\bigcup\limits_{t = 1}^{T_\text{simulation}} \mathcal{N}^{(t)} = \mathcal{N}$ since $\mathcal{N}^{(t)}$ is disjoint.}} $\mathcal{N}^{(t)}$ and $\mathcal{U}^{(t)}$} (i.e., partitions).
    \item Train the classifier using {a randomly sampled subset of} this dataset (or the \textit{learning} set). {The size of this subset is given by $\vert\mathcal{N}^{(t)}\setminus\mathcal{U}^{(t)}\vert := \lceil (1 - q_\text{explotation})N_\text{UE}^{(t)} \rceil$, where $q_\text{exploitation}$ is a fraction of the total dataset size and $N_\text{UE}^{(t)}$ is the stratified subset size (UE count).}
    \item Use this classifier to predict the band switch for the {set of} UEs {belonging to the subset $\mathcal{U}^{(t)}$}.
\end{itemize}

\textbf{Classifier choice:} For the classifier we \review{consider} two options: DNN and extreme gradient boosting (XGBoost) trees.  {XGBoost is a scalable tree boosting system that achieves high classification performance \cite{XGBoost_Chen16}}. DNN is a feed-forward architecture that uses layers of neurons of a given depth $d$ and width $w$ {and can approximate arbitrary functions under assumptions on the activation functions \cite{Deep_Goodfellow16}}. An activation function defines the output of a neuron with respect to its inputs.  A DNN optimizes a convex loss function through a learning rate $\eta > 0 $.   XGBoost optimizes an objective function containing a convex loss function (e.g., binary logistic loss) and a regularization term $\alpha \Vert\mathbf{w} \Vert_1+\frac{1}{2}\lambda \Vert \mathbf{w} \Vert_2^2 + \gamma T$, where $\mathbf{w}$ is the vector containing the leaf weights in the boosted tree, $\alpha$ and $\lambda$ are the regularization terms for their respective norms, $\gamma$ is the complexity control, and $T$ is the number of leaves.

\begin{algorithm}[!t]
    \caption{Measurement gap free band switch policy}
    \label{alg:algorithm}
     \DontPrintSemicolon
     \KwIn{Parameters listed in Tables~\ref{table:rf_parameters_ifho}~and~\ref{table:hyperparams_isho}.  Instantaneous rates by UE location per frequency band (or technology) of a given BS association area.  Simulation time $T_\text{simulation}$. Set of all UEs $\mathcal{N}$ and target UEs $\mathcal{U}\subset\mathcal{N}$, defined by $q_\text{exploitation}$.}
    \KwOut{A vector $\mathbf{\hat y}$ containing a prediction whether the measurement gap-free band switch should be granted or denied for the set $\mathcal{U}$ in the same BS association area, a confusion matrix $\mathbf{C}$, and the area under the receiver operating characteristic (ROC) curve for the set.}
\For{$t := 1$ \KwTo $T_{\text{\rm simulation}}$} {
         {Stratified sampling with no replacement from $\mathcal{N}$ into $\mathcal{N}^{(t)}$.  Let $N_\text{UE}^{(t)} := \vert\mathcal{N}^{(t)}\vert$.}\;
         $N_\text{learning}^{(t)} := \lceil (1 - q_\text{exploitation})\cdot N_\text{UE}^{(t)}\rceil$\;
        At random, sample with no replacement a subset of users from $\mathcal{N}^{(t)}$ {into a set $\mathcal{U}^{(t)}$.}\;
        Build the learning dataset $[\mathbf{X} \,\vert\, \mathbf{y}]$ for UEs $\{1,2,\ldots, N_\text{learning}^{(t)}\}$, where $\mathbf{X}$ is in Table~\ref{tab:features} and $\mathbf{y}$ is based on \eqref{Eq:LegVar2}, both for all $N_\text{learning}^{(t)}$ UEs. \;
        Randomly split the data $[\mathbf{X} \,\vert\, \mathbf{y}]$ into a training and a test data (using $q_\text{training}$ split ratio). \;
        Train the DNN classifier {balancing the weights of classes $\mathbf{y}$} in the training data and use grid search on $K$-fold cross-validation to tune the hyperparameters based on the binary cross-entropy loss function \cite{Deep_Goodfellow16}.\;
        
        \ForAll {$u \in \mathcal{U}^{(t)}$}{%
            Use the DNN classifier to predict $\hat y^{(u)}$ based on $\mathbf{X}^{(u)}$.\;
        } 
        
        Obtain $\xi$ the area under the ROC curve for this model using $\hat{\mathbf{y}} := [\hat y^{(u)}]$.\;
        Build the confusion matrix $\mathbf{C}$ by observing $\mathbf{y}$ and $\hat{\mathbf{y}}$.\;
        {Invalidate the DNN classifier.}\;
    }
\end{algorithm}

\textbf{Classifier training:} We train the hyperparameters of the classifiers using grid search and $K$-fold cross-validation.  The list of learning features is shown in Table~\ref{tab:features}.  Let the feature matrix be denoted by $\mathbf{X}\in\mathbb{R}^{N_\text{UE}\times p}, p > 2$. The industry standards require two features for the band switch decision, as we showed in Section~\ref{sec:band_change_policies}.  The supervisory label vector is a column vector $\mathbf{y}\in\{0,1\}^{N_\text{UE}}$, where 0 means the band switch was denied and 1 means granted based on \eqref{Eq:LegVar1}.

\begin{figure*}[!t]
\centering
\begin{tikzpicture}[node distance=4.5cm, auto, scale=0.95]
\tikzstyle{every node}=[font=\small]
\linespread{0.9}
\node[text width=2cm] at (0.5,0.2) {Requests};
\draw[line width=1.5, ->, >=latex] (-0.5,0) -- (11,0) node[below] {time $t$};

\foreach \x in {2.25,3.5,5,6.4,7.5,8.5,9.5}
    \draw[->, >=latex] (\x,0) -- ++(0,0.5);

\foreach \x/\y in {2.25/grant,3.5/deny,5/deny,6.4/grant,7.5/grant,8.5/deny,9.5/grant} {
    \draw[->, >=latex] (\x+0.5,-1) -- ++(0,0.5) node[above] {\smaller \y};
    \draw[fill=gray!30] (\x,-1) rectangle ++(0.5,0.25);
};

\node[text width=2cm] at (0.5,-0.8) {Legacy};
\draw[line width=1.5, ->, >=latex] (-0.5,-1) -- (11,-1) node[below] {};
\node[text width=2cm] at (0.5,-1.8) {Proposed};
\draw[line width=1.5, ->, >=latex] (-0.5,-2) -- (11,-2) node[below] {};

\draw[dashed] (1.5,-2.5) -- ++(0,3.5);
\draw[dashed] (4.75,-2.5) -- ++(0,3.5);

\draw[line width=1.5, <->, >=latex] (1.5,1) -- (4.75,1) node[midway,above] {Learning};
\draw[line width=1.5, <->, >=latex] (4.75,1) -- (10.25,1) node[midway,above] {Exploitation Phase};

\draw[->, >=latex] (5.75,-2) -- ++(0,0.5) node[above] {\smaller grant}  node[below,yshift=-0.75cm,xshift=0.1cm] {\smaller \textsf{bad}};
\draw[->, >=latex] (6.75,-2) -- ++(0,0.5) node[above] {\smaller grant} node[below, text width=5em, align=center, yshift=-0.75cm,xshift=1cm] {\smaller \textsf{good \\ predictions}};
\draw[->, >=latex] (7.75,-2) -- ++(0,0.5) node[above] {\smaller grant};

\draw[->, >=latex] (8.75,-2) -- ++(0,0.5) node[above] {\smaller deny};
\draw[->, >=latex] (9.75,-2) -- ++(0,0.5) node[above] {\smaller deny} node[below,yshift=-0.75cm,xshift=0.1cm] {\smaller \textsf{bad}};

\end{tikzpicture}
\vspace*{-1em}\caption{Proposed band switch time diagram.  The shaded gray rectangles represent the transmission gaps.}
\label{fig:proposed_diagram}
\end{figure*}
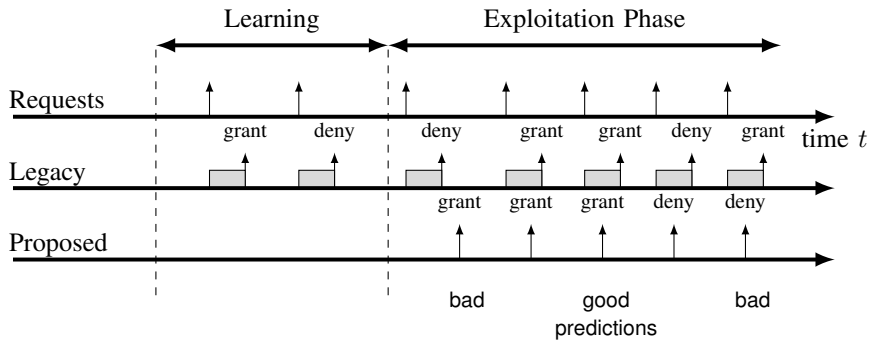

Our proposed approach is shown in comparison to the legacy approach in Fig.~\ref{fig:proposed_diagram} and it operates in two phases: a) learning phase and b) exploitation phase. 

\textbf{Learning phase:} In the learning phase, the UE follows the legacy approach discussed earlier while the proposed algorithm stores the learning features $\mathbf{X}$ and $\mathbf{y}$.  Machine learning is then applied on this data to build a classifier that estimates band switch decisions but without the need for a measurement gap.  During this phase, we let all UEs request band switches by setting $\mathbf{x}_7$ to unity (or $r_\text{threshold}$ to $+\infty$).  This is in order for the classifier to learn the relationship between channels regardless of band switch requests.  We use both DNN and XGBoost classification algorithms in the implementation of this phase {and compute class weights for imbalanced the classes $\mathbf{y}$.}

\textbf{Exploitation phase:} In the exploitation phase, the classifier uses prediction to eliminate the measurement gap for the set of UEs $\mathcal{U}$ which were not used in the learning phase.  The predicted decision either grants or denies the band switch from the $j^\text{th}$ band.  The exploitation phase essentially represents the \textit{generalization capacity} of the classifier {for the current radio frame}. 

{\textbf{Classifier invalidation:} the invalidation (i.e., purging and retraining) of classifiers dealing with wireless channels in online-learning setting prevents changes in the channel state information from not being reflected onto the classifier \cite{8665922}.  Given that we only allow the band switching to take place in the beginning of the radio frame as stated earlier, this leaves the classifier with $(T_\text{RF} - 1)$ timeslots to realistically collect data and train.  The number of measurements per radio frame is  $\vert\mathcal{N}\setminus \mathcal{U}\vert = \lceil (1 - q_\text{exploitation}) \cdot \vert\mathcal{N}\vert \rceil$, after which the classifier must be invalidated. 
}

\section{Performance Measures}\label{sec:performance}

In this section we describe our choices of the performance measures to benchmark our algorithm.  These measures describe the performance of both the QoE of the users and the performance of the classifier.

\subsection{Effective Achievable Rate}
{For the different policies discussed,} we evaluate the effective rate of all the users in the network using \eqref{eq:ear2}. We are interested in the statistics of the effective achievable rates. Namely, the cumulative distribution function (CDF) and the mean.

\subsection{Confusion Matrix}
We define the misclassification count $E\colon 0 \le E \le n$ as the number of incorrectly predicted band switches during the exploitation phase.  We build a confusion matrix $\mathbf{C} \in\mathbb{Z}_+^{2\times 2}$ having the true and predicted band switch decision counts and write
\begin{equation}
    E := \mathrm{Tr} (\mathbf{J}\mathbf{C}^\top),
    \label{eq:misclassification_error}
\end{equation}
where $\mathbf{J}$ is a $2\times 2$ anti-diagonal identity matrix.  The misclassification error $\mu$ can be derived by dividing $E$ by $n:=\lfloor q_\text{exploitation}N_\text{UE} \rfloor$. 

\subsection{ROC Area Under the Curve}
The receiver operating characteristic (ROC) curve is a two-dimensional curve used to visualize classifiers based on the tradeoff between hit rates and false alarm rates.  To compare the performance of classifiers, we reduce the ROC performance to a single scalar quantity known as the ROC area under the curve \cite{An_Fawcett06}.  This area $\xi\colon 0 \le \xi \le 1.0$ where 0.5 means the classifier is as good as a random guess and 1.0 means it produces perfect prediction.

So far, we have discussed the different band switch policies, including our proposed policy. We have highlighted the desired performance measures we are interested in. In the next section, we discuss the data we use and how we construct in detail.

\section{Dataset Construction}\label{sec:dataset}

To test the performance of the proposed algorithm, we rely on the DeepMIMO dataset \cite{DeepMIMO_Alkhateeb19}. The choice of this dataset is based on its use of accurate ray-tracing tools to generate spatially and spectrally correlated channels for specific scenarios. Hence, we avoid using oversimplified mathematical models that could lead to misleading results. A better choice would be to use a dataset that is based on actual field-measurements. However, such dataset is not available yet to the best of our knowledge, \review{and is highly non-trivial to generate.}

\begin{figure}[!t]
    \centering
    \includegraphics[scale=0.45]{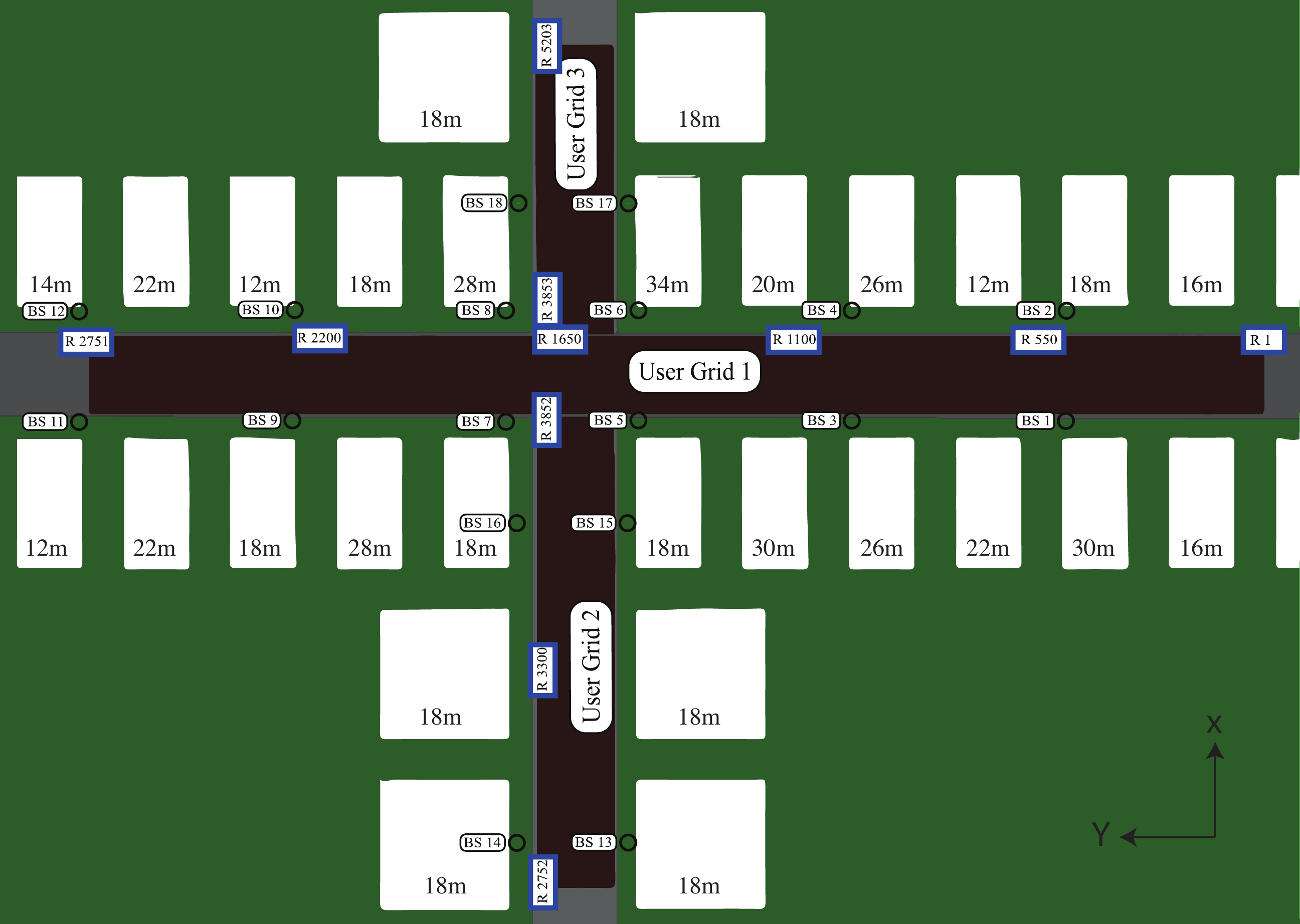}
    \vspace*{-1em}
    \caption{Scenario O1 of the DeepMIMO dataset \cite{DeepMIMO_Alkhateeb19}.  We use base station (BS) 3 and users on User Grid 1.}
    \label{fig:deepmimo}
\end{figure}

In the O1 outdoor scenario of the DeepMIMO dataset, the UEs are placed on a uniform grid on a main street for both the sub-6 GHz and mmWave frequency bands, where the BS uses OFDM and uniform planar array (UPA) antennas. The adopted O1 scenario is shown in Fig.~\ref{fig:deepmimo}.

\subsection{Channel Coherence Time}
The channel coherence time over which the channel remains highly correlated is known to be given by  \cite{Fundamentals_Ghosh10}
\begin{equation}\label{Eq:TypCohTi}
    T_C(\alpha)\approx\frac{c}{f_c v_s \sin \alpha},
\end{equation}
where $c$ is the speed of light, $v_s$ is the speed of the UE, $\alpha$ is the angle between the direction of travel and the direction of the BS,  $v_s \sin \alpha$ is the relative speed of the user with regards to the BS, and $f_c$ is the center frequency. This equation has been widely used to measure the channel coherence time for the sub-6 GHz range, where omni-directional antennas are used. However, at mmWave, where directional antennas along with beamforming are employed to combat the high isotropic path loss, \eqref{Eq:TypCohTi} does not accurately measure coherence time \cite{The_Va17}. This is because by combining directional antenna arrays with beamforming, the signal power is focused on a beamwidth-defined angular space directed towards the UE location. Hence, only the variations in the channel within this angular space are relevant, which increases the channel coherence compared to \eqref{Eq:TypCohTi}. The coherence time of the beamformed channel, referred to as the beam coherence time, is given by \cite{The_Va17}
\begin{equation}
    T_C(\alpha) \approx\frac{D}{v_s\sin\alpha}\frac{\Theta}{2},
    \label{eq:cohmmwave}
\end{equation}
where $D$ is the Euclidean distance from the serving BS and $\Theta$ is the beamwidth of the beams used by serving BS (in radians).  Since UEs are located at different locations with different distances to the BS, they have different coherence times. However, to maintain a fixed frame length for all users connected at the same band, we assume the cell-center beam coherence time (e.g., the $1^\text{st}$ percentile). This conservative assumption is also motivated by the practical case where the BS may not have full knowledge of the UE parameters, such as their distance and accurate location.
To sum up, we assume that the coherence time for sub-6 GHz and mmWave bands is given by
\begin{align}
    T_C^\text{sub-6} &:=\left(\frac{c}{f_c v_s \sin \alpha}\right)_{0.01} ,
    \label{eq:cohsub6}\\
T_C^\text{mmWave} &:= \left( \frac{D}{v_s\sin\alpha}\frac{\Theta}{2} \right)_{0.01},\label{eq:cohmm}
\end{align}
respectively, where $(\mathcal{X})_{0.01}$ is the $1^\text{st}$ percentile of the set $\mathcal{X}$. For the frame time, we set the frame duration to be equal to the channel coherence time for simplicity. Hence, the overheads for beam training and band switch, mentioned earlier, are related to a single parameter, which is the coherence time.

\subsection{Band-Selective Blockage}
Further, we choose to have occasional blockage in the mmWave frequency band.  To generate this blockage using DeepMIMO, we generate two channels: one with blockage and the other without blockage.  We further combine the mmWave channels into one by introducing a Bernoulli random variable for the $i^\text{th}$ UE:
\begin{align}
    b_i &\sim \textrm{Bernoulli}(p), \qquad i = 1, 2, \ldots, N_\text{UE} \label{eq:blockage} \\
    \mathbf{h}^{(i)}[t] &= b_i\mathbf{h}_b^{(i)} [t] + (1 - b_i) \mathbf{h}_{nb}^{(i)}[t],
\end{align}
where $p$ is the blockage probability, $\mathbf{h}_b^{(i)}$ is the channel with blockage on the first mmWave path, and $\mathbf{h}_{nb}^{(i)}$ is the channel with no blockage, all for the $i^\text{th}$ UE. Hence, some locations along the street are assumed to be blocked from the BS (non-line of sight), while others have a line of sight.  {To study the behavior of the proposed algorithm against uncertainty, we vary the blockage probability $p$ in the simulation during the exploitation phase.}

\subsection{Analog Beamforming}

We adopt a multi-antenna setup, where the BS employs a UPA of $M_y^{(j)}$ and $M_z^{(j)}$ antennas in the elevation and azimuth directions respectively at the $j^\text{th}$ band.  Therefore, we write the channel in \eqref{eq:channel} as $\mathbf{h}\in\mathbb{C}^{M_y^{(j)} M_z^{(j)} \times 1}$ in a vectorized form. In our implementation of analog beamforming, we focus on discrete Fourier transform (DFT) codebooks. We focus on DFT codebooks {as they are a common practice for effective codebook design when the channels are spatially correlated \cite{1214832}}.  Let the $M \times N_\mathrm{CB}^{(j)}$ matrix $\bF^{(j)}$ be the concatenation of $M$ beamforming vectors in the codebook $\mathcal{F}^{(j)}$, then the matrix $\bF^{(j)}$ is constructed as%
\begin{equation}
    \bF^{(j)} = \bF^{(j)}_z \kron \bF^{(j)}_y
\end{equation}
where $\bF^{(j)}_y \in \mathbb{C}^{M^{(j)}_y \times M^{(j)}_y}$ and $\bF^{(j)}_z\in\mathbb{C}^{M^{(j)}_z \times M^{(j)}_z}$ concatenate the DFT codebook vectors in the y and z directions for the $j^\text{th}$ frequency band. In the next section, we use this dataset to evaluate the performance of the proposed algorithm and compare it with the other algorithms discussed in Section~\ref{sec:band_change_policies}.%

\section{Simulation Results}\label{sec:results}

\Rev{In this section, we evaluate the proposed algorithm using the DeepMIMO dataset described in Section~\ref{sec:dataset} using the performance measures outined in Section~\ref{sec:performance}.}

\subsection{Setup}

\begin{table}[!t]
\setlength\doublerulesep{0.5pt}
\caption{Radio environment parameters}
\vspace*{-1em}
\label{table:rf_parameters_ifho}
\centering
\begin{tabular}{ lr } 
\hhline{==}
Parameter & Value \\
 \hline
Subcarrier bandwidth (sub-6, mmWave) & $(180, 1800)$ kHz \\
Center frequency & (3.5, 28) GHz \\
UE noise figure & 7 dB \\
DeepMIMO Scenario O1 Base Station & 3 \\
DeepMIMO Scenario O1 number of antennas $(M_x,M_y,M_z)$ & $(1,64,4)$\\
DeepMIMO Scenario O1 OFDM limit & 64 \\
Band switch threshold rate for sub-6 GHz $r_\text{threshold}^\text{sub-6}$ & 1.72 Mbps\\
Band switch threshold rate for mmWave $r_\text{threshold}^\text{mmWave}$ & 7.00 Mbps\\
Measurement gap fraction of coherence time $\rho$ & 0.6 \\
\hhline{==}
\end{tabular}
\end{table}

\begin{table}[!t]
    \setlength\doublerulesep{0.5pt}
    \caption{Hyperparameters of classifiers used in the implementation of our algorithm}
    \label{table:hyperparams_isho}
    \vspace*{-1em}
    \centering
    \begin{tabular}{ p{0.35\linewidth}r|p{0.35\linewidth}r } 
    \hhline{====}
    \multicolumn{2}{c|}{DNN} & \multicolumn{2}{c}{XGBoost} \\
    \hline
    Parameter & Value & Parameter & Value \\
    \hline
    Exploitation split $q_\text{exploitation}$ & 0.8 & Exploitation split $q_\text{exploitation}$ & 0.8 \\
    $K$-fold cross-validation $K$ & 2 & $K$-fold cross-validation $K$ & 2 \\
    Optimizer & \cite{KingmaB14} & $\ell_1$ regularization term $\alpha$ & \{0,1\} \\
    Learning rate $\eta$ & 0.05 & $\ell_2$ regularization term $\lambda$ & \{0,1\} \\
    Activation function & sigmoid & Complexity control term $\gamma$ & \{0,0.02,0.04\} \\ 
    Depth of neural network $d$ & \{1,3,5\} & Sample weights & \{0.5,0.7\} \\
    Width of the hidden layer $w$ & \{3,5,10\} & Child weights & \{0,10\} \\
    \hhline{====}
    \end{tabular}
\end{table}

The DNN and XGBoost classifier hyperparameters are both shown in Table~\ref{table:hyperparams_isho}.  
As mentioned earlier, the users are placed on a uniform grid on a main street in the association area of this co-sited BS such that the $i^\text{th}$ UE has the Cartesian coordinate $(x_i, y_i, z_i)$. This grid has an area of 550 m and a width of 35 m for an area of 19,250 square meters.  The spacing between every two adjacent users in this uniform grid is 0.2 m. The height of all UEs $z_i = 2$ m is constant throughout the simulation.  {We set the mmWave channel blockage probability in \eqref{eq:blockage} to $p = 0.4$ during the \textit{learning} phase.  Given that we choose not to perform oversampling in beamforming, $N_\text{CB} := M_y$.} We set the beam training time $T_\text{beam} := 1~\mu\text{s}$ \cite{The_Va17}.   {Further, in \eqref{eq:cohsub6} and \eqref{eq:cohmm}, we set $\alpha \sim \text{Uniform}(0,\pi)$ and $\Theta := 102 / M_y$  following \cite{The_Va17}.}
The users move at a vehicular speed $v_s = 50$ km/h within the BS association area every discrete time step $t$. There is a total of $54,\! 480$ users in the simulation {divided into partitions of $5,\! 448$ users, hence $T_\text{simulation} = 10$ radio frames.} In an attempt to find the {\textit{absolute}} training dataset size that best maximizes the ROC area, we choose from $q_\text{training} \in \{1\times 10^{-3},5\times 10^{-3},7\times 10^{-3},1\times 10^{-2},3\times 10^{-2},5\times 10^{-2},7\times 10^{-2},1\times 10^{-1},3\times 10^{-1},4\times 10^{-1},5\times 10^{-1},7\times 10^{-1}\}$.
With $12$ OFDM subcarriers per physical resource block (PRB) and a subcarrier spacing of $15$ kHz, we have the bandwidth $B = 180~\text{kHz}$ per PRB.  We use one PRB for the sub-6 GHz frequency band and ten PRBs for the mmWave frequency band.  In other words, $B^\text{sub-6} = 180~\text{kHz}$ and $B^\text{mmWave} = 1800$~kHz. \Rev{Initially,} the coherence times for sub-6 GHz and mmWave based on \eqref{eq:cohsub6} and \eqref{eq:cohmm} are 6.17 and 19.16~ms respectively.  {This is justified due to the increased beamforming gain at mmWave, which slows down the time fluctuation of the channel \cite{The_Va17, 1343892}}.  We choose the transmit energy of $0.1$ W/Hz at mmWave and set the transmit energy at sub-6 GHz to $1$ W/Hz.  \Rev{We set the measurement gap fraction $\rho$ to 0.6 aligned with the industry standards of the gap duration per frame \cite{3gpp36133}.} Further, we set the band switch thresholds as 1.72 Mbps and 7.00 Mbps for sub-6 GHz and mmWave {as the empirical means of the throughput distributions}. The exploitation ratio $q_\text{exploitation}$ of 0.8 means that the {total} number of UEs that will use the trained model for band switching is $0.8\times 54,\!480 = 43,\!584$ UEs.  In the learning phase, we set the band switch request threshold to $+\infty$. This allows us to capture all the available spatial correlation information between the channels without any omission. 

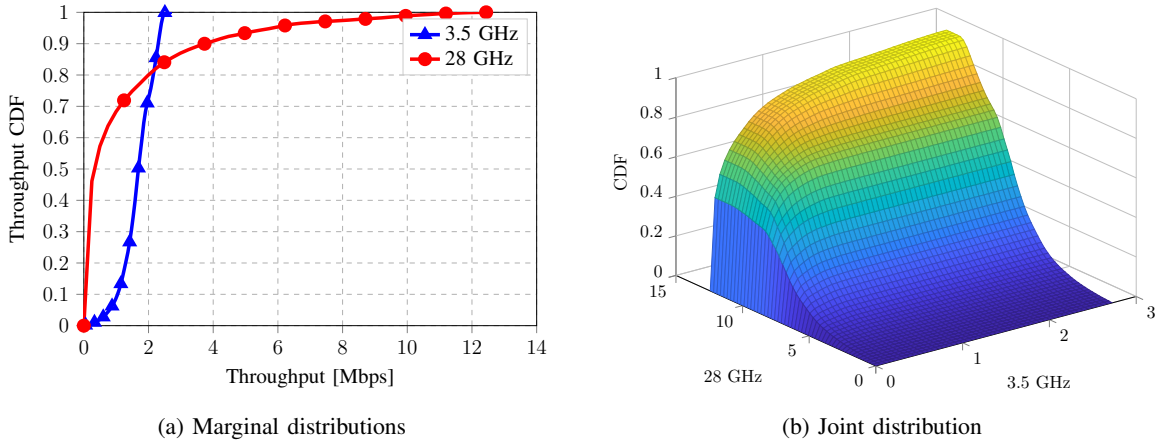
\begin{figure}[!t]
    \centering
  \subfloat[Marginal distributions]{\resizebox{0.45\textwidth}{!}{
\begin{tikzpicture}

\begin{axis}[
width=4.2in,
height=3.1in,
legend cell align={left},
legend entries={{3.5 GHz},{28 GHz}},
legend style={draw=white!80.0!black},
tick align=outside,
tick pos=left,
x grid style={white!69.01960784313725!black, dashed},
xlabel={Throughput [Mbps]},
xmajorgrids,
xmin=0, xmax=14,
xtick={0,2,...,14},
y grid style={white!69.01960784313725!black, dashed},
ylabel={Throughput CDF},
ymajorgrids,
ymin=0,
ymax=1,
ytick={0,0.1,0.2,...,1},
]

\addplot [line width=2pt, blue, mark=triangle, mark size=3, mark repeat=5]
table [row sep=\\]{
0.0563249621099911	0.00151431718061674 \\
0.110762394911422	0.00270741556534508 \\
0.165199827712853	0.00399229074889868 \\
0.219637260514283	0.00591960352422908 \\
0.274074693315714	0.00816813509544787 \\
0.328512126117145	0.0104854992657856 \\
0.382949558918576	0.0128258076358297 \\
0.437386991720006	0.0161756607929515 \\
0.491824424521437	0.0191813509544787 \\
0.546261857322868	0.0227147577092511 \\
0.600699290124298	0.0276248164464024 \\
0.655136722925729	0.0333838105726872 \\
0.70957415572716	0.0396016886930984 \\
0.764011588528591	0.046140785609398 \\
0.818449021330021	0.0538959251101322 \\
0.872886454131452	0.0623852790014684 \\
0.927323886932883	0.072503671071953 \\
0.981761319734314	0.0841363803230543 \\
1.03619875253574	0.0979717327459619 \\
1.09063618533718	0.114285058737151 \\
1.14507361813861	0.133764684287812 \\
1.19951105094004	0.156364720998532 \\
1.25394848374147	0.181626284875184 \\
1.3083859165429	0.207622063142438 \\
1.36282334934433	0.234260279001468 \\
1.41726078214576	0.267001651982379 \\
1.47169821494719	0.305020190895742 \\
1.52613564774862	0.348086453744493 \\
1.58057308055005	0.394938509544787 \\
1.63501051335148	0.448375550660793 \\
1.68944794615291	0.502890969162996 \\
1.74388537895434	0.55036251835536 \\
1.79832281175577	0.596939243759178 \\
1.85276024455721	0.647003487518355 \\
1.90719767735864	0.684218979441997 \\
1.96163511016007	0.710375367107195 \\
2.0160725429615	0.732378854625551 \\
2.07050997576293	0.756791483113069 \\
2.12494740856436	0.786963105726872 \\
2.17938484136579	0.819268538913363 \\
2.23382227416722	0.854992657856094 \\
2.28825970696865	0.891336270190896 \\
2.34269713977008	0.929125367107195 \\
2.39713457257151	0.964574155653451 \\
2.45157200537294	0.989766886930984 \\
2.50600943817437	0.998990455212922 \\
2.56044687097581	0.999449339207048 \\
2.61488430377724	0.999724669603524 \\
2.66932173657867	0.999908223201175 \\
2.7237591693801	1 \\
};
\addplot [line width=2pt, red, mark=*, mark size=3, mark repeat=5, mark options={solid}]
table [row sep=\\]{%
0 0 \\
0.248764983698804	0.461384911894273 \\
0.497529967232926	0.572687224669604 \\
0.746294950767049	0.637366923641704 \\
0.995059934301171	0.682842327459618 \\
1.24382491783529	0.719025330396476 \\
1.49258990136942	0.74825624082232 \\
1.74135488490354	0.773426027900147 \\
1.99011986843766	0.799421806167401 \\
2.23888485197178	0.821562958883994 \\
2.4876498355059	0.840675477239354 \\
2.73641481904003	0.855680983847284 \\
2.98517980257415	0.869539280469897 \\
3.23394478610827	0.880781938325991 \\
3.48270976964239	0.890602055800294 \\
3.73147475317652	0.899665014684288 \\
3.98023973671064	0.907580763582966 \\
4.22900472024476	0.915748898678414 \\
4.47776970377888	0.923113986784141 \\
4.726534687313	0.927909324522761 \\
4.97529967084713	0.933576541850221 \\
5.22406465438125	0.93880781938326 \\
5.47282963791537	0.943419603524229 \\
5.72159462144949	0.94771016886931 \\
5.97035960498362	0.952826725403818 \\
6.21912458851774	0.957966226138033 \\
6.46788957205186	0.96227973568282 \\
6.71665455558598	0.965216593245228 \\
6.9654195391201	0.967075073421439 \\
7.21418452265423	0.969300660792952 \\
7.46294950618835	0.970906754772394 \\
7.71171448972247	0.972168685756241 \\
7.96047947325659	0.97372889133627 \\
8.20924445679072	0.975334985315712 \\
8.45800944032484	0.977262298091043 \\
8.70677442385896	0.979029001468429 \\
8.95553940739308	0.980726872246696 \\
9.20430439092721	0.982287077826726 \\
9.45306937446133	0.984489720998532 \\
9.70183435799545	0.986623531571219 \\
9.95059934152957	0.988573788546256 \\
10.1993643250637	0.990340491923642 \\
10.4481293085978	0.991831864904552 \\
10.6968942921319	0.99327734948605 \\
10.9456592756661	0.994952276064611 \\
11.1944242592002	0.99607654185022 \\
11.4431892427343	0.99720080763583 \\
11.6919542262684	0.998370961820852 \\
11.9407192098026	0.99912812041116 \\
12.1894841933367	0.999678781204112 \\
12.4382491768708	1 \\
};




\end{axis}

\end{tikzpicture}}}%
    \hfil
    \subfloat[Joint distribution]{\resizebox{0.45\textwidth}{!}{\includegraphics{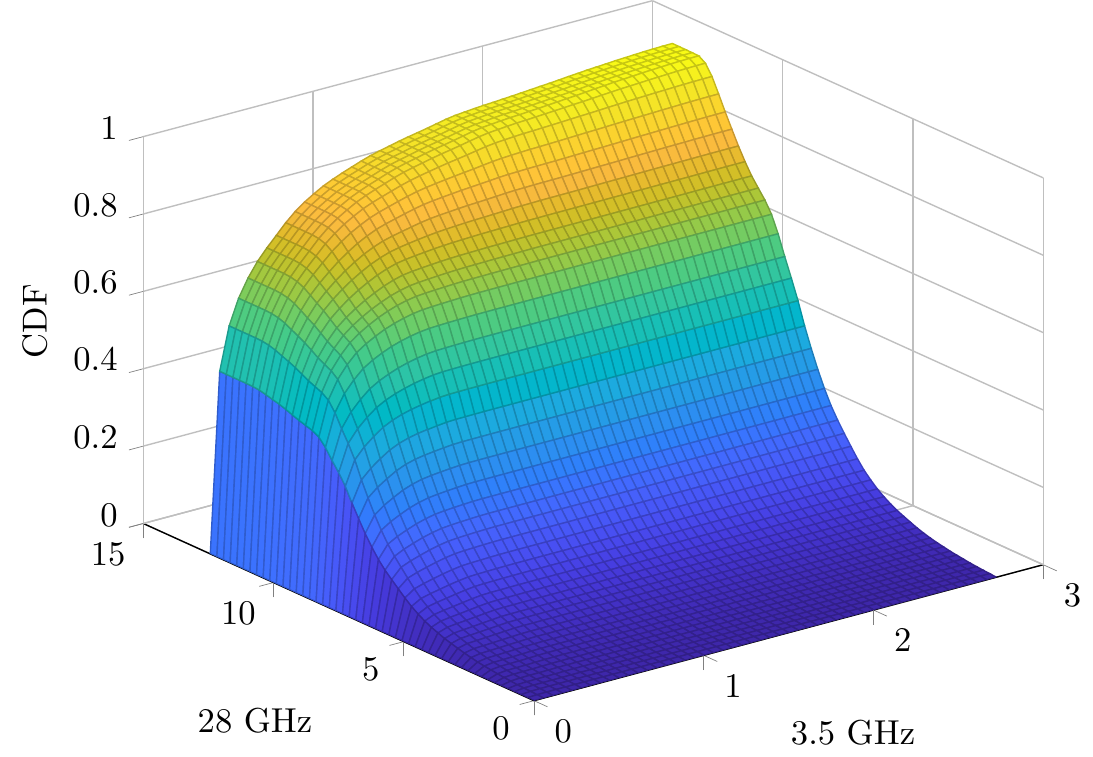}}}%
    \caption{Instantaneous throughput distributions for both sub-6 and mmWave frequency bands.} 
    \label{fig:pdf}
\end{figure}

We simulate the radio environment (given in Table~\ref{table:rf_parameters_ifho}) using three different scenarios: 

\begin{itemize}
    \item \textit{Scenario A:} All users start in sub-6 and attempt to change band to mmWave.
    \item \textit{Scenario B:} All users start in mmWave and attempt to change band to sub-6 GHz.
    \item \textit{Scenario C:} 70\% are in sub-6 and 30\% are in mmWave.
\end{itemize}

We refer to the source code for the details of the implementation of this simulation \cite{mycode}.  Before presenting the results for these scenarios, we start with analyzing the raw data from the dataset. In Fig.~\ref{fig:pdf}, we show the distribution of the effective throughput over all users for the $3.5$ GHz and the $28$ GHz bands. From the marginal distributions, we can see that the effective throughput for the mmWave bands goes up to 12~Mbps, while it only goes up to 3~Mbps for the $3.5$~GHz band. This is due to the large bandwidth that is available in the mmWave band. However, due to blockage, these high rates only occur with a small probability, since the two CDF curves cross at 0.8. Overall, the figure shows that the effective rate at mmWave can be very high, due to the large bandwidth, or very low due to the blockage. This wide range of rates motivates the optimized design of the band switch policy, since ineffective design can cause significant deterioration in the UE throughput, or can prevent the user from harvesting a high rate from mmWave bands.

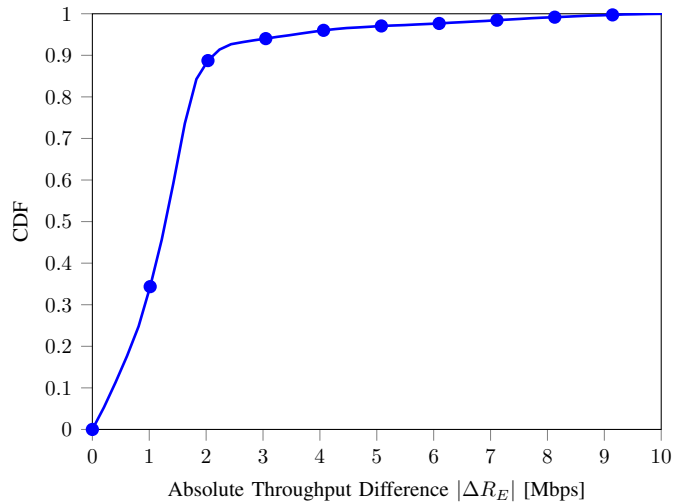
\begin{figure}[!t]
    \centering
	\scalebox{1}{
\begin{tikzpicture}
\tikzstyle{every node}=[font=\smaller,scale=0.8]

\begin{axis}[
width=3.6in,
height=2.8in,
tick align=outside,
tick pos=left,
xlabel={Absolute Throughput Difference $\vert \Delta R_E \vert$ [Mbps]},
xmin=0, xmax=10,
xtick={0,1,2,...,10},
ylabel={CDF},
ymin=0,ymax=1,
ytick={0,0.1,...,1.1},
]

\addplot [line width=1pt, blue, mark=*, mark size=2, mark repeat=5, mark options={solid}]
table [row sep=\\]{%
0 0 \\
0.20338819291052	0.0530699339207048 \\
0.406587904986487	0.112931350954479 \\
0.609787617062454	0.176739170337739 \\
0.812987329138422	0.247980910425844 \\
1.01618704121439	0.34356644640235 \\
1.21938675329036	0.456383076358297 \\
1.42258646536632	0.591478524229075 \\
1.62578617744229	0.736187591776799 \\
1.82898588951826	0.842740455212922 \\
2.03218560159423	0.887343979441997 \\
2.23538531367019	0.913867474302496 \\
2.43858502574616	0.926785058737151 \\
2.64178473782213	0.931809838472834 \\
2.8449844498981	0.936238069016153 \\
3.04818416197406	0.940207415565345 \\
3.25138387405003	0.94447503671072 \\
3.454583586126	0.94814610866373 \\
3.65778329820197	0.952321953010279 \\
3.86098301027793	0.956291299559471 \\
4.0641827223539	0.960122980910426 \\
4.26738243442987	0.963105726872247 \\
4.47058214650584	0.965652533039648 \\
4.6737818585818	0.967120961820852 \\
4.87698157065777	0.968772944199707 \\
5.08018128273374	0.970654368575624 \\
5.28338099480971	0.971755690161528 \\
5.48658070688567	0.972742290748899 \\
5.68978041896164	0.974164831130691 \\
5.89298013103761	0.975403817914831 \\
6.09617984311357	0.976688693098385 \\
6.29937955518954	0.978455396475771 \\
6.50257926726551	0.979809104258444 \\
6.70577897934148	0.981300477239354 \\
6.90897869141744	0.982677129221733 \\
7.11217840349341	0.984466776798825 \\
7.31537811556938	0.986049926578561 \\
7.51857782764535	0.987610132158591 \\
7.72177753972131	0.989330947136564 \\
7.92497725179728	0.990684654919237 \\
8.12817696387325	0.99178597650514 \\
8.33137667594922	0.992956130690162 \\
8.53457638802518	0.994286894273128 \\
8.73777610010115	0.995525881057269 \\
8.94097581217712	0.996237151248165 \\
9.14417552425309	0.997269640234949 \\
9.34737523632905	0.998370961820852 \\
9.55057494840502	0.998806901615272 \\
9.75377466048099	0.999472283406755 \\
9.95697437255695	0.99981644640235 \\
10.1601740846329	1 \\
};




\end{axis}

\end{tikzpicture}}%
	\caption{The distribution of the absolute difference between the rate on 3.5 GHz and 28 GHz.}
	\label{fig:Diff}
\end{figure}

From the joint distribution, the general trend is that a higher throughput on $3.5$ GHz means a higher throughput on $28$ GHz. This is due to the correlation between the channels caused by common propagation paths, reflectors, and obstructions. By simple computations using the joint distributions, one can see that the $25\%$ of the users can get higher throughput on the $28$ GHz band. Intuitively, these are the users who do not suffer from blockage and are at a short distance from the BS. These users would benefit from operating at mmWave. To quantify this gain, we plot Fig.~\ref{fig:Diff}, which shows the distribution of the absolute difference between the throughput at $3.5$ GHz and $28$ GHz, \review{$\vert \Delta R_E \vert := \vert R_E^\text{sub-6} - R_E^\text{mmWave} \vert $}.  Next, we analyze the performance of the different band switch policies discussed earlier.

\subsection{Band Switch Policies}
\subsubsection{Scenario A}
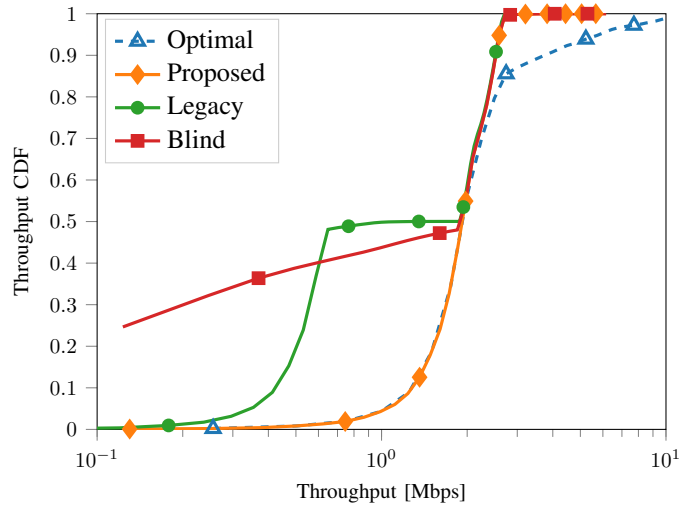
\begin{figure}[!t]
	\centering
	\scalebox{1}{
\begin{tikzpicture}
\tikzstyle{every node}=[font=\smaller,scale=0.8]
\definecolor{color0}{rgb}{0.12156862745098,0.466666666666667,0.705882352941177}
\definecolor{color1}{rgb}{1,0.498039215686275,0.0549019607843137}
\definecolor{color2}{rgb}{0.172549019607843,0.627450980392157,0.172549019607843}
\definecolor{color3}{rgb}{0.83921568627451,0.152941176470588,0.156862745098039}

\begin{semilogxaxis}[
width=3.6in,
height=2.8in,
legend cell align={left},
legend entries={{Optimal},{Proposed},{Legacy},{Blind}},
legend style={at={(0.32,0.64)}, anchor=south east, draw=white!80.0!black, nodes={scale=1.25, transform shape}},
tick align=outside,
tick pos=left,
x grid style={white!69.01960784313725!black,dashed},
xlabel={Throughput [Mbps]},
xmin=0.1, xmax=10,
y grid style={white!69.01960784313725!black,dashed},
ylabel={Throughput CDF},
ymin=0, ymax=1, 
ytick={0,0.1,0.2,...,1}
]
\addplot [color0, line width=1.2pt, dashed, mark=triangle, mark size=3,mark repeat=10, mark options={solid}]
table [row sep=\\]{%
0 0 \\
0.25575835003606	0.0020879221732746 \\
0.504380611808197	0.00835168869309838 \\
0.753002873580334	0.0197549559471366 \\
1.00162513535247	0.0439151982378855 \\
1.25024739712461	0.089665932452276 \\
1.49886965889675	0.186215124816446 \\
1.74749192066888	0.341960352422907 \\
1.99611418244102	0.560916850220264 \\
2.24473644421316	0.698169052863436 \\
2.4933587059853	0.79662261380323 \\
2.74198096775743	0.855382709251101 \\
2.99060322952957	0.869860499265786 \\
3.23922549130171	0.880988436123348 \\
3.48784775307385	0.890739720998531 \\
3.73647001484598	0.899825624082232 \\
3.98509227661812	0.907695484581498 \\
4.23371453839026	0.915932452276065 \\
4.4823368001624	0.923205763582966 \\
4.73095906193453	0.927955212922173 \\
4.97958132370667	0.933668318649045 \\
5.22820358547881	0.93880781938326 \\
5.47682584725095	0.943442547723935 \\
5.72544810902308	0.947756057268722 \\
5.97407037079522	0.952918502202643 \\
6.22269263256736	0.958012114537445 \\
6.4713148943395	0.962325624082232 \\
6.71993715611163	0.96526248164464 \\
6.96855941788377	0.967143906020558 \\
7.21718167965591	0.969323604992658 \\
7.46580394142805	0.9709296989721 \\
7.71442620320018	0.972191629955947 \\
7.96304846497232	0.97372889133627 \\
8.21167072674446	0.975357929515419 \\
8.4602929885166	0.977285242290749 \\
8.70891525028873	0.979029001468429 \\
8.95753751206087	0.980726872246696 \\
9.20615977383301	0.982310022026432 \\
9.45478203560515	0.984489720998532 \\
9.70340429737728	0.986646475770925 \\
9.95202655914942	0.988573788546256 \\
10.2006488209216	0.990340491923642 \\
10.4492710826937	0.991831864904552 \\
10.6978933444658	0.99327734948605 \\
10.946515606238	0.994952276064611 \\
11.1951378680101	0.99607654185022 \\
11.4437601297822	0.99720080763583 \\
11.6923823915544	0.998370961820852 \\
11.9410046533265	0.99912812041116 \\
12.1896269150987	0.999678781204112 \\
12.4382491768708	1 \\
};
\addplot [line width=1.2pt, color1, mark=diamond*, mark size=3, mark phase=1, mark repeat=5, mark options={solid}]
table [row sep=\\]{%
0.130157679763093	0.000825991189427313 \\
0.253179271262263	0.00204203377386197 \\
0.376200862761433	0.00417584434654919 \\
0.499222454260604	0.00819107929515418 \\
0.622244045759774	0.0131470264317181 \\
0.745265637258944	0.0189289647577092 \\
0.868287228758115	0.0292309104258443 \\
0.991308820257285	0.042859765051395 \\
1.11433041175646	0.0601596916299559 \\
1.23735200325563	0.0862243024963289 \\
1.3603735947548	0.12543593979442 \\
1.48339518625397	0.179010646108664 \\
1.60641677775314	0.241258259911894 \\
1.72943836925231	0.327184287812041 \\
1.85245996075148	0.436398678414097 \\
1.97548155225065	0.54923825256975 \\
2.09850314374982	0.657580763582966 \\
2.22152473524899	0.717373348017621 \\
2.34454632674816	0.781227055800294 \\
2.46756791824733	0.861004038179148 \\
2.5905895097465	0.947962555066079 \\
2.71361110124567	0.995089941262849 \\
2.83663269274484	0.997292584434655 \\
2.95965428424401	0.998233296622614 \\
3.08267587574318	0.998829845814978 \\
3.20569746724235	0.999059287812041 \\
3.32871905874152	0.999174008810572 \\
3.45174065024069	0.999426395007342 \\
3.57476224173986	0.999587004405286 \\
3.69778383323903	0.999701725403818 \\
3.8208054247382	0.999770558002936 \\
3.94382701623737	0.999839390602055 \\
4.06684860773655	0.999862334801762 \\
4.18987019923572	0.999862334801762 \\
4.31289179073489	0.999908223201174 \\
4.43591338223406	0.999908223201174 \\
4.55893497373323	0.999908223201174 \\
4.6819565652324	0.999954111600587 \\
4.80497815673157	0.999954111600587 \\
4.92799974823074	0.999977055800293 \\
5.05102133972991	0.999977055800293 \\
5.17404293122908	0.999977055800293 \\
5.29706452272825	0.999977055800293 \\
5.42008611422742	0.999977055800293 \\
5.54310770572659	0.999977055800293 \\
5.66612929722576	0.999977055800293 \\
5.78915088872493	0.999977055800293 \\
5.9121724802241	0.999977055800293 \\
6.03519407172327	0.999977055800293 \\
6.15821566322244	1 \\
};
\addplot [line width=1.2pt, color2, mark=*, mark size=2, mark phase=2, mark repeat=10, mark options={solid}]
table [row sep=\\]{%
0.0610517216399932	0.00126193098384728 \\
0.119744788505776	0.00360223935389134 \\
0.178437855371558	0.0094988986784141 \\
0.23713092223734	0.0173228707782672 \\
0.295823989103123	0.0314335535976505 \\
0.354517055968905	0.0529552129221733 \\
0.413210122834687	0.0890923274596182 \\
0.471903189700469	0.152877202643172 \\
0.530596256566252	0.239009728340675 \\
0.589289323432034	0.370273494860499 \\
0.647982390297816	0.481116923641703 \\
0.706675457163599	0.485659875183553 \\
0.765368524029381	0.488688509544787 \\
0.824061590895163	0.492061306901615 \\
0.882754657760946	0.494562224669603 \\
0.941447724626728	0.49694842143906 \\
1.00014079149251	0.498370961820851 \\
1.05883385835829	0.499151064610866 \\
1.11752692522408	0.499449339207048 \\
1.17621999208986	0.499724669603524 \\
1.23491305895564	0.499931167400881 \\
1.29360612582142	0.500091776798825 \\
1.3522991926872	0.500160609397944 \\
1.41099225955299	0.50018355359765 \\
1.46968532641877	0.500229441997063 \\
1.52837839328455	0.500229441997063 \\
1.58707146015033	0.500275330396476 \\
1.64576452701612	0.500275330396476 \\
1.7044575938819	0.500298274596182 \\
1.76315066074768	0.500298274596182 \\
1.82184372761346	0.500298274596182 \\
1.88053679447925	0.500298274596182 \\
1.93922986134503	0.534852239353891 \\
1.99792292821081	0.585100036710719 \\
2.05661599507659	0.638720631424376 \\
2.11530906194237	0.681282121879589 \\
2.17400212880816	0.710054148311307 \\
2.23269519567394	0.733778450807635 \\
2.29138826253972	0.760760829662261 \\
2.3500813294055	0.793685756240822 \\
2.40877439627129	0.829593428781204 \\
2.46746746313707	0.868323237885462 \\
2.52616053000285	0.908544419970631 \\
2.58485359686863	0.949591593245227 \\
2.64354666373442	0.980864537444933 \\
2.7022397306002	0.998623348017621 \\
2.76093279746598	0.999380506607929 \\
2.81962586433176	0.999678781204111 \\
2.87831893119754	0.999908223201174 \\
2.93701199806333	1 \\
};
\addplot [line width=1.2pt, color3, mark=square*, mark size=2, mark phase=3, mark repeat=10, mark options={solid}]
table [row sep=\\]{%
0.123164313425837	0.246443649045521 \\
0.246328626686992	0.322044787077827 \\
0.369492939948148	0.363757342143906 \\
0.492657253209303	0.387435756240822 \\
0.615821566470458	0.403267254038179 \\
0.738985879731613	0.415817731277533 \\
0.862150192992768	0.426257342143906 \\
0.985314506253924	0.436398678414097 \\
1.10847881951508	0.445920521292217 \\
1.23164313277623	0.454226321585903 \\
1.35480744603739	0.461247246696035 \\
1.47797175929854	0.467143906020558 \\
1.6011360725597	0.472191629955947 \\
1.72430038582085	0.476298641703378 \\
1.84746469908201	0.479809104258444 \\
1.97062901234317	0.545200073421439 \\
2.09379332560432	0.654368575624082 \\
2.21695763886548	0.715652533039648 \\
2.34012195212663	0.778496696035242 \\
2.46328626538779	0.858365455212922 \\
2.58645057864894	0.945232195301028 \\
2.7096148919101	0.99502110866373 \\
2.83277920517125	0.997292584434655 \\
2.95594351843241	0.998233296622614 \\
3.07910783169356	0.998829845814978 \\
3.20227214495472	0.999059287812041 \\
3.32543645821587	0.999174008810573 \\
3.44860077147703	0.999426395007342 \\
3.57176508473818	0.999587004405287 \\
3.69492939799934	0.999701725403818 \\
3.81809371126049	0.999770558002937 \\
3.94125802452165	0.999839390602056 \\
4.0644223377828	0.999862334801762 \\
4.18758665104396	0.999862334801762 \\
4.31075096430511	0.999908223201175 \\
4.43391527756627	0.999908223201175 \\
4.55707959082743	0.999908223201175 \\
4.68024390408858	0.999954111600587 \\
4.80340821734973	0.999954111600587 \\
4.92657253061089	0.999977055800294 \\
5.04973684387205	0.999977055800294 \\
5.1729011571332	0.999977055800294 \\
5.29606547039436	0.999977055800294 \\
5.41922978365551	0.999977055800294 \\
5.54239409691667	0.999977055800294 \\
5.66555841017782	0.999977055800294 \\
5.78872272343898	0.999977055800294 \\
5.91188703670013	0.999977055800294 \\
6.03505134996129	0.999977055800294 \\
6.15821566322244	1 \\
};




\end{semilogxaxis}

\end{tikzpicture}}%
	\caption{The distribution of the effective throughput for Scenario A under different band switch polices.}
	\label{fig:Sc_a}
\end{figure} 

We start with Scenario A, where all the users start in $3.5$ GHz band and can request a band switch to the $28$ GHz band. The results are shown in Fig.~\ref{fig:Sc_a}, where we show the distribution of the effective throughput under different band switch policies. Starting with the legacy approach, the effect of the measurement gap is clear in the figure and results in a performance gap compared to the optimal policy, especially in the low rate regime, where the negative impact of the measurement gap is detrimental. Precisely, the effect is more dominant for users suffering from a low throughput at sub-6 GHz and requested a band switch from the BS, but their request got denied because the throughput at mmWave is also low, possibly due to blockage. Hence, having a measurement gap makes their throughput even worse. This can be observed from the curve for the small throughput regime.  Further, we observe that there are more users in the small throughput range (i.e., less than $0.5$ Mbps) in the blind policy than the legacy policy. This can be justified as follows: for the same class of users suffering from extremely low throughput in the sub-6 GHz band, but would not benefit from switching to the mmWave, the legacy policy instructs them to stay in the sub-6 GHz band at the expense of a measurement gap, while the blind policy switches these users to the mmWave band, which deteriorates their throughput even more. However, there is a point where the throughput at mmWave is around the same as the sub-6 GHz. At this point, the blind policy is more efficient since it has the advantage of not requiring a measurement gap. 

For the proposed algorithm, Fig.~\ref{fig:Sc_a} also shows that it has the best performance compared to the previous two; it is identical to the optimal in the low rate regime and identical to the other policies in the high rate regime. This is due to: (\textit{i}) the elimination of the measurement gap, hence users with low throughput do not suffer more if their band switch request got denied as in the legacy approach, and (\textit{ii}) the accurate band switch decisions, which prevents switching users to a band with low throughput as in the blind policy. Note that there is a performance gap between the three policies and the optimal in the high rate regime. This performance gap is due to the band switch threshold introduced in these policies, but missing from the optimal. Hence, users with high throughput in the sub-6 GHz band do not benefit from the higher throughput in the mmWave bands following these policies. But this is not the case for the optimal algorithm, since the BS picks the band with the maximum throughput each frame without a threshold. However, as we will show later, our proposed algorithm can overcome this issue by increasing the band switch threshold, without losing its accuracy. Finally, to quantify the gains provided by the different band switch policies, we list the mean effective throughput in Table \ref{table:results}. Based on the values for Scenario A, the proposed algorithm provides a gain {in the mean effective throughput} of $39\%$ and $37\%$ over the blind and the legacy policies, respectively, and just $20\%$ behind the optimal algorithm. Overall, the results for this scenario are promising and show the effectiveness of the proposed algorithm.

\begin{figure}[!t]
	\centering
	\scalebox{1}{
\begin{tikzpicture}
\tikzstyle{every node}=[font=\smaller,scale=0.8]
\definecolor{color0}{rgb}{0.12156862745098,0.466666666666667,0.705882352941177}
\definecolor{color1}{rgb}{1,0.498039215686275,0.0549019607843137}
\definecolor{color2}{rgb}{0.172549019607843,0.627450980392157,0.172549019607843}
\definecolor{color3}{rgb}{0.83921568627451,0.152941176470588,0.156862745098039}

\begin{semilogxaxis}[
width=3.6in,
height=2.8in,
legend cell align={left},
legend entries={{Optimal},{Proposed},{Legacy},{Blind}},
legend style={at={(0.99,0.03)}, anchor=south east, draw=white!80.0!black, nodes={scale=1.25, transform shape}},
tick align=outside,
tick pos=left,
x grid style={white!69.01960784313725!black,dashed},
xlabel={Throughput [Mbps]},
xmin=0.1, xmax=12,
y grid style={white!69.01960784313725!black,dashed},
ylabel={Throughput CDF},
ymin=0, ymax=1, 
ytick={0,0.1,0.2,...,1}
]
\addplot [color0, line width=1.2pt, dashed, mark=triangle, mark size=3,mark repeat=10, mark phase=5, mark options={solid}]
table [row sep=\\]{%
0.251923681032548	0.00426762114537445 \\
0.500624201355778	0.0158544419970631 \\
0.749324721679007	0.0371696035242291 \\
0.998025242002237	0.0762894640234949 \\
1.24672576232547	0.16322503671072 \\
1.4954262826487	0.305570851688693 \\
1.74412680297193	0.521888766519824 \\
1.99282732329515	0.676486784140969 \\
2.24152784361838	0.767529368575624 \\
2.49022836394161	0.840147760646109 \\
2.73892888426484	0.855841593245228 \\
2.98762940458807	0.869745778267254 \\
3.2363299249113	0.880827826725404 \\
3.48503044523453	0.890693832599119 \\
3.73373096555776	0.8997109030837 \\
3.98243148588099	0.907603707782673 \\
4.23113200620422	0.915794787077827 \\
4.47983252652745	0.923159875183554 \\
4.72853304685068	0.927909324522761 \\
4.97723356717391	0.933668318649046 \\
5.22593408749714	0.93880781938326 \\
5.47463460782037	0.943419603524229 \\
5.7233351281436	0.947733113069016 \\
5.97203564846683	0.952849669603524 \\
6.22073616879006	0.957966226138032 \\
6.46943668911329	0.962302679882526 \\
6.71813720943652	0.965239537444934 \\
6.96683772975975	0.967098017621145 \\
7.21553825008298	0.969323604992658 \\
7.46423877040621	0.9709296989721 \\
7.71293929072943	0.972168685756241 \\
7.96163981105266	0.97372889133627 \\
8.21034033137589	0.975357929515418 \\
8.45904085169912	0.977262298091042 \\
8.70774137202235	0.979029001468429 \\
8.95644189234558	0.980726872246696 \\
9.20514241266881	0.982287077826725 \\
9.45384293299204	0.984489720998531 \\
9.70254345331527	0.986623531571219 \\
9.9512439736385	0.988573788546255 \\
10.1999444939617	0.990340491923641 \\
10.448645014285	0.991831864904552 \\
10.6973455346082	0.99327734948605 \\
10.9460460549314	0.994952276064611 \\
11.1947465752546	0.99607654185022 \\
11.4434470955779	0.99720080763583 \\
11.6921476159011	0.998370961820852 \\
11.9408481362243	0.99912812041116 \\
12.1895486565476	0.999678781204112 \\
12.4382491768708	1 \\
};
\addplot [line width=1.2pt, color1, mark=diamond*, mark size=3, mark phase=1, mark repeat=10, mark options={solid}]
table [row sep=\\]{%
0.251923681032548	0.00426762114537445 \\
0.500624201355778	0.0158544419970631 \\
0.749324721679007	0.0371696035242291 \\
0.998025242002237	0.0762894640234949 \\
1.24672576232547	0.16322503671072 \\
1.4954262826487	0.305639684287812 \\
1.74412680297193	0.521980543318649 \\
1.99282732329515	0.676739170337739 \\
2.24152784361838	0.767758810572687 \\
2.49022836394161	0.840147760646109 \\
2.73892888426484	0.855841593245228 \\
2.98762940458807	0.869745778267254 \\
3.2363299249113	0.880827826725404 \\
3.48503044523453	0.890693832599119 \\
3.73373096555776	0.8997109030837 \\
3.98243148588099	0.907603707782673 \\
4.23113200620422	0.915794787077827 \\
4.47983252652745	0.923159875183554 \\
4.72853304685068	0.927909324522761 \\
4.97723356717391	0.933668318649046 \\
5.22593408749714	0.93880781938326 \\
5.47463460782037	0.943419603524229 \\
5.7233351281436	0.947733113069016 \\
5.97203564846683	0.952849669603524 \\
6.22073616879006	0.957966226138032 \\
6.46943668911329	0.962302679882526 \\
6.71813720943652	0.965239537444934 \\
6.96683772975975	0.967098017621145 \\
7.21553825008298	0.969323604992658 \\
7.46423877040621	0.9709296989721 \\
7.71293929072943	0.972168685756241 \\
7.96163981105266	0.97372889133627 \\
8.21034033137589	0.975357929515418 \\
8.45904085169912	0.977262298091043 \\
8.70774137202235	0.979029001468429 \\
8.95644189234558	0.980726872246696 \\
9.20514241266881	0.982287077826725 \\
9.45384293299204	0.984489720998531 \\
9.70254345331527	0.986623531571219 \\
9.9512439736385	0.988573788546255 \\
10.1999444939617	0.990340491923642 \\
10.448645014285	0.991831864904552 \\
10.6973455346082	0.99327734948605 \\
10.9460460549314	0.994952276064611 \\
11.1947465752546	0.99607654185022 \\
11.4434470955779	0.99720080763583 \\
11.6921476159011	0.998370961820852 \\
11.9408481362243	0.99912812041116 \\
12.1895486565476	0.999678781204112 \\
12.4382491768708	1 \\
};
\addplot [line width=1.2pt, color2, mark=*, mark size=2, mark phase=3, mark repeat=5, mark options={solid}]
table [row sep=\\]{%
0.249736244856994	0.048481093979442 \\
0.498481406734826	0.464757709251101 \\
0.747226568612659	0.836683186490455 \\
0.995971730490492	0.880988436123348 \\
1.24471689236832	0.91017345814978 \\
1.49346205424616	0.930892070484581 \\
1.74220721612399	0.946746512481645 \\
1.99095237800182	0.962348568281938 \\
2.23969753987965	0.962623898678414 \\
2.48844270175749	0.962623898678414 \\
2.73718786363532	0.962623898678414 \\
2.98593302551315	0.962623898678414 \\
3.23467818739099	0.962623898678414 \\
3.48342334926882	0.962623898678414 \\
3.73216851114665	0.962623898678414 \\
3.98091367302448	0.962623898678414 \\
4.22965883490232	0.962623898678414 \\
4.47840399678015	0.962623898678414 \\
4.72714915865798	0.962623898678414 \\
4.97589432053582	0.962623898678414 \\
5.22463948241365	0.962623898678414 \\
5.47338464429148	0.962623898678414 \\
5.72212980616931	0.962623898678414 \\
5.97087496804715	0.962623898678414 \\
6.21962012992498	0.962623898678414 \\
6.46836529180281	0.962623898678414 \\
6.71711045368064	0.965216593245228 \\
6.96585561555848	0.967075073421439 \\
7.21460077743631	0.969300660792952 \\
7.46334593931414	0.970906754772394 \\
7.71209110119197	0.972168685756241 \\
7.96083626306981	0.97372889133627 \\
8.20958142494764	0.975357929515419 \\
8.45832658682547	0.977262298091043 \\
8.7070717487033	0.979029001468429 \\
8.95581691058114	0.980726872246696 \\
9.20456207245897	0.982287077826725 \\
9.4533072343368	0.984489720998531 \\
9.70205239621463	0.986623531571219 \\
9.95079755809247	0.988573788546255 \\
10.1995427199703	0.990340491923642 \\
10.4482878818481	0.991831864904552 \\
10.697033043726	0.99327734948605 \\
10.9457782056038	0.994952276064611 \\
11.1945233674816	0.99607654185022 \\
11.4432685293595	0.99720080763583 \\
11.6920136912373	0.998370961820852 \\
11.9407588531151	0.99912812041116 \\
12.189504014993	0.999678781204112 \\
12.4382491768708	1 \\
};
\addplot [line width=1.2pt, color3, mark=square*, mark size=2, mark phase=1, mark repeat=15, mark options={solid}]
table [row sep=\\]{%
0.250614762259805	0.00718153450807636 \\
0.49934199521105	0.0196861233480176 \\
0.748069228162294	0.0440758076358297 \\
0.996796461113539	0.0874403450807636 \\
1.24552369406478	0.177565161527166 \\
1.49425092701603	0.322228340675477 \\
1.74297815996727	0.549444750367107 \\
1.99170539291852	0.722512848751836 \\
2.24043262586976	0.849508994126285 \\
2.48915985882101	0.961499632892805 \\
2.73788709177225	0.962623898678414 \\
2.9866143247235	0.962623898678414 \\
3.23534155767474	0.962623898678414 \\
3.48406879062599	0.962623898678414 \\
3.73279602357723	0.962623898678414 \\
3.98152325652848	0.962623898678414 \\
4.23025048947972	0.962623898678414 \\
4.47897772243097	0.962623898678414 \\
4.72770495538221	0.962623898678414 \\
4.97643218833345	0.962623898678414 \\
5.2251594212847	0.962623898678414 \\
5.47388665423594	0.962623898678414 \\
5.72261388718719	0.962623898678414 \\
5.97134112013843	0.962623898678414 \\
6.22006835308968	0.962623898678414 \\
6.46879558604092	0.962623898678414 \\
6.71752281899217	0.965216593245228 \\
6.96625005194341	0.967075073421439 \\
7.21497728489466	0.969300660792952 \\
7.4637045178459	0.970906754772394 \\
7.71243175079715	0.972168685756241 \\
7.96115898374839	0.97372889133627 \\
8.20988621669964	0.975357929515419 \\
8.45861344965088	0.977262298091043 \\
8.70734068260212	0.979029001468429 \\
8.95606791555337	0.980726872246696 \\
9.20479514850461	0.982287077826725 \\
9.45352238145586	0.984489720998532 \\
9.7022496144071	0.986623531571219 \\
9.95097684735835	0.988573788546255 \\
10.1997040803096	0.990340491923642 \\
10.4484313132608	0.991831864904552 \\
10.6971585462121	0.99327734948605 \\
10.9458857791633	0.994952276064611 \\
11.1946130121146	0.99607654185022 \\
11.4433402450658	0.997200807635829 \\
11.6920674780171	0.998370961820851 \\
11.9407947109683	0.99912812041116 \\
12.1895219439195	0.999678781204111 \\
12.4382491768708	1 \\
};




\end{semilogxaxis}

\end{tikzpicture}}%
	\caption{The distribution of the effective throughput for Scenario B under different band switch polices.}
	\label{fig:Sc_b}
\end{figure}
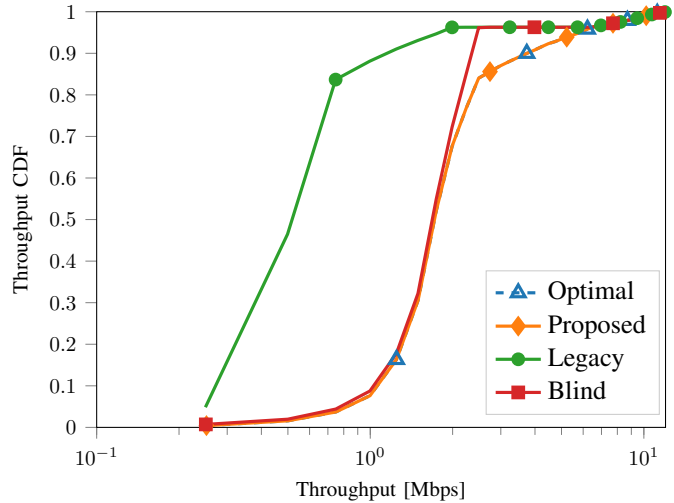 

\begin{table}[!t]
    \setlength\doublerulesep{0.5pt}
    \caption{ \review{Normalized} mean effective throughput for different scenarios}
    \label{table:results}
    \vspace*{-1em}
    \centering
    \resizebox{0.5\textwidth}{!}{%
        \begin{tabular}{ lc@{\hskip 0.35in}c@{\hskip 0.35in}c@{\hskip 0.35in}c} 
        \hhline{=====}
         & \multicolumn{4}{c}{Normalized mean effective throughput $R_E$} \\ 
         \cline{2-5} 
         Scenario & Legacy & Blind & Proposed & Optimal  \\
        \hline
        Scenario A & 0.55 & 0.54 & 0.75  & 1.00 \\ 
        Scenario B & 0.43 & 0.88 & 1.00 & 1.00 \\
        Scenario C & 0.35 & 0.76 & 1.00 & 1.00 \\

        \hhline{=====}
        \end{tabular}%
    }
\end{table}

\subsubsection{Scenario B} For the second scenario, the results are shown in Fig.~\ref{fig:Sc_b}. There are a few differences between this scenario and Scenario A. Firstly, all policies achieve the same performance in the high rate regime, which is due to the assumption that all users start in the mmWave band. To be precise, the high throughput regime (above 4~Mbps) can only be achieved on mmWave bands as shown in Fig.~\ref{fig:pdf}. Hence, given that the users start in the mmWave band and this range is above the band switch threshold, these users remain in mmWave regardless of the policy, which justifies the identical performance of the different policies in the high throughput regime. Secondly, the blind {policy} achieves an identical performance to the optimal {policy} for the low throughput region. To justify this, we need the data in Table~\ref{table:ho_req_granted}, which show the number of band switch requests and the number of the granted ones for each policy and each scenario. From this table, we can see that up to around $70\%$ of the band switch requests in this scenario are granted (assuming the optimal policy).  Hence, the blind policy is identical to the optimal policy $70\%$ of the time. Among this $70\%$ of the users are the users who suffer from extremely low throughput at mmWave, mostly due to blockage. Hence, the blind algorithm results in the optimal decision for these users which justifies the identical performance for the low throughput regime. However, the blind policy also makes the wrong decision $30\%$ of the time, which results in a gap between this policy and the optimal policy in the medium throughput regime. Finally, the legacy is always the worst in this scenario, which is due the measurement gap and the fact that the blind  {policy} is accurate $70\%$ of the time without having a measurement gap.  The averages of the effective throughput are also shown in Table~\ref{table:results}. The proposed policy achieves $130\%$ {gain in the throughput} compared to the legacy policy and $13\%$ compared to the blind policy. Also, the effective throughput for the proposed is almost identical to the optimal policy. 

\begin{table}[!t]
    \setlength\doublerulesep{0.5pt}
    \caption{Band switch grants and requests based on the band switch policy and the scenario} 
    \label{table:ho_req_granted}
    \vspace*{-1em}
    \centering
        \begin{tabular}{l|lcc} 
        \hhline{====}
         Scenario & Policy & Band switches requested & Band switches granted  \\
        
       \hline
        Scenario A & Legacy & 22,458 & 2,751 \\ 
                   & Blind & 22,458 & 22,458 \\ 
                   & Proposed & 22,458 &  2,724  \\ 
        \hline
        Scenario B & Legacy & 41,609 & 32,558 \\ 
                   & Blind & 41,609 & 41,609 \\ 
                   & Proposed & 41,609 & 32,569 \\ 
        \hline
        Scenario C & Legacy & 43,033  &  21,514 \\ 
                   & Blind & 43,033 & 43,033 \\ 
                   & Proposed & 43,033  & 21,619 \\ 
        \hhline{====}
        \end{tabular}%
\end{table}

\subsubsection{Scenario C} 
\begin{figure}[!t]
	\centering
	\scalebox{1}{
\begin{tikzpicture}
\tikzstyle{every node}=[font=\smaller,scale=0.8]
\definecolor{color0}{rgb}{0.12156862745098,0.466666666666667,0.705882352941177}
\definecolor{color1}{rgb}{1,0.498039215686275,0.0549019607843137}
\definecolor{color2}{rgb}{0.172549019607843,0.627450980392157,0.172549019607843}
\definecolor{color3}{rgb}{0.83921568627451,0.152941176470588,0.156862745098039}

\begin{semilogxaxis}[%
width=3.6in,
height=2.8in,
legend cell align={left},
legend entries={{Optimal},{Proposed},{Legacy},{Blind}},
legend style={at={(0.99,0.02)}, anchor=south east, draw=white!80.0!black, nodes={scale=1.25, transform shape}},
tick align=outside,
tick pos=left,
x grid style={white!69.01960784313725!black,dashed},
xlabel={Throughput [Mbps]},
xmin=0.1, xmax=11,
y grid style={white!69.01960784313725!black,dashed},
ylabel={Throughput CDF},
ymin=0, ymax=1, 
ytick={0,0.1,0.2,...,1}
]
\addplot [color0, line width=1.2pt, dashed, mark=triangle, mark size=3,mark repeat=10, mark phase=5, mark options={solid}]
table [row sep=\\]{%
0 0 \\
0.251923681032548	0.00752569750367107 \\
0.500624201355778	0.0263169970631424 \\
0.749324721679007	0.0570851688693098 \\
0.998025242002237	0.121351872246696 \\
1.24672576232547	0.239376835535976 \\
1.4954262826487	0.441538179148311 \\
1.74412680297193	0.637963472834068 \\
1.99282732329515	0.725587371512482 \\
2.24152784361838	0.814519089574156 \\
2.49022836394161	0.840813142437592 \\
2.73892888426484	0.855841593245228 \\
2.98762940458807	0.869745778267254 \\
3.2363299249113	0.880827826725404 \\
3.48503044523453	0.890693832599119 \\
3.73373096555776	0.8997109030837 \\
3.98243148588099	0.907603707782673 \\
4.23113200620422	0.915794787077827 \\
4.47983252652745	0.923159875183554 \\
4.72853304685068	0.927909324522761 \\
4.97723356717391	0.933668318649046 \\
5.22593408749714	0.93880781938326 \\
5.47463460782037	0.943419603524229 \\
5.7233351281436	0.947733113069016 \\
5.97203564846683	0.952849669603524 \\
6.22073616879006	0.957966226138032 \\
6.46943668911329	0.962302679882526 \\
6.71813720943652	0.965239537444934 \\
6.96683772975975	0.967098017621145 \\
7.21553825008298	0.969323604992658 \\
7.46423877040621	0.9709296989721 \\
7.71293929072943	0.972168685756241 \\
7.96163981105266	0.97372889133627 \\
8.21034033137589	0.975357929515418 \\
8.45904085169912	0.977262298091042 \\
8.70774137202235	0.979029001468429 \\
8.95644189234558	0.980726872246696 \\
9.20514241266881	0.982287077826725 \\
9.45384293299204	0.984489720998531 \\
9.70254345331527	0.986623531571219 \\
9.9512439736385	0.988573788546255 \\
10.1999444939617	0.990340491923641 \\
10.448645014285	0.991831864904552 \\
10.6973455346082	0.99327734948605 \\
10.9460460549314	0.994952276064611 \\
11.1947465752546	0.99607654185022 \\
11.4434470955779	0.99720080763583 \\
11.6921476159011	0.998370961820852 \\
11.9408481362243	0.99912812041116 \\
12.1895486565476	0.999678781204112 \\
12.4382491768708	1 \\
};
\addplot [line width=1.2pt, color1, mark=diamond*, mark size=3, mark phase=1, mark repeat=10, mark options={solid}]
table [row sep=\\]{%
0 0 \\
0.251923681032548	0.00752569750367107 \\
0.500624201355778	0.0263169970631424 \\
0.749324721679007	0.0570851688693098 \\
0.998025242002237	0.121420704845815 \\
1.24672576232547	0.239422723935389 \\
1.4954262826487	0.441675844346549 \\
1.74412680297193	0.638078193832599 \\
1.99282732329515	0.725816813509545 \\
2.24152784361838	0.814656754772394 \\
2.49022836394161	0.840813142437592 \\
2.73892888426484	0.855841593245227 \\
2.98762940458807	0.869745778267254 \\
3.2363299249113	0.880827826725404 \\
3.48503044523453	0.890693832599119 \\
3.73373096555776	0.8997109030837 \\
3.98243148588099	0.907603707782672 \\
4.23113200620422	0.915794787077827 \\
4.47983252652745	0.923159875183554 \\
4.72853304685068	0.927909324522761 \\
4.97723356717391	0.933668318649046 \\
5.22593408749714	0.93880781938326 \\
5.47463460782037	0.943419603524229 \\
5.7233351281436	0.947733113069016 \\
5.97203564846683	0.952849669603524 \\
6.22073616879006	0.957966226138032 \\
6.46943668911329	0.962302679882525 \\
6.71813720943652	0.965239537444934 \\
6.96683772975975	0.967098017621145 \\
7.21553825008298	0.969323604992658 \\
7.46423877040621	0.9709296989721 \\
7.71293929072943	0.972168685756241 \\
7.96163981105266	0.97372889133627 \\
8.21034033137589	0.975357929515418 \\
8.45904085169912	0.977262298091042 \\
8.70774137202235	0.979029001468429 \\
8.95644189234558	0.980726872246696 \\
9.20514241266881	0.982287077826725 \\
9.45384293299204	0.984489720998531 \\
9.70254345331527	0.986623531571219 \\
9.9512439736385	0.988573788546255 \\
10.1999444939617	0.990340491923641 \\
10.448645014285	0.991831864904552 \\
10.6973455346082	0.99327734948605 \\
10.9460460549314	0.994952276064611 \\
11.1947465752546	0.99607654185022 \\
11.4434470955779	0.99720080763583 \\
11.6921476159011	0.998370961820852 \\
11.9408481362243	0.99912812041116 \\
12.1895486565476	0.999678781204111 \\
12.4382491768708	1 \\
};
\addplot [line width=1.2pt, color2, mark=*, mark size=2, mark phase=2, mark repeat=10, mark options={solid}]
table [row sep=\\]{%
0 0 \\
0.143260004847471	0.0212233847283407 \\
0.285528926715781	0.0884498898678414 \\
0.427797848584091	0.304263032305433 \\
0.570066770452401	0.650789280469897 \\
0.712335692320711	0.802358663729809 \\
0.854604614189022	0.843772944199706 \\
0.996873536057332	0.865386380323054 \\
1.13914245792564	0.881057268722467 \\
1.28141137979395	0.894594346549192 \\
1.42368030166226	0.907259544787078 \\
1.56594922353057	0.91707966226138 \\
1.70821814539888	0.925316629955947 \\
1.85048706726719	0.932681718061674 \\
1.9927559891355	0.939358480176211 \\
2.13502491100381	0.947090675477239 \\
2.27729383287212	0.952620227606461 \\
2.41956275474043	0.955557085168869 \\
2.56183167660874	0.958333333333333 \\
2.70410059847705	0.960742474302496 \\
2.84636952034536	0.963610499265786 \\
2.98863844221367	0.967763399412629 \\
3.13090736408198	0.971526248164464 \\
3.27317628595029	0.976046255506608 \\
3.4154452078186	0.980222099853157 \\
3.55771412968691	0.983778450807636 \\
3.69998305155522	0.986715308370044 \\
3.84225197342353	0.989491556534508 \\
3.98452089529184	0.991808920704846 \\
4.12678981716015	0.993369126284875 \\
4.26905873902846	0.994332782672541 \\
4.41132766089677	0.994791666666667 \\
4.55359658276508	0.995089941262849 \\
4.6958655046334	0.995296439060206 \\
4.83813442650171	0.9956406020558 \\
4.98040334837002	0.99607654185022 \\
5.12267227023833	0.996397760646109 \\
5.26494119210664	0.996810756240822 \\
5.40721011397495	0.997063142437592 \\
5.54947903584326	0.997292584434655 \\
5.69174795771157	0.997659691629956 \\
5.83401687957988	0.997957966226138 \\
5.97628580144819	0.998187408223201 \\
6.1185547233165	0.998370961820852 \\
6.26082364518481	0.998738069016153 \\
6.40309256705312	0.99894456681351 \\
6.54536148892143	0.999357562408223 \\
6.68763041078974	0.999609948604993 \\
6.82989933265805	0.99981644640235 \\
6.97216825452636	0.999954111600587 \\
7.11443717639467	1 \\
};
\addplot [line width=1.2pt, color3, mark=square*, mark size=2, mark phase=2, mark repeat=15, mark options={solid}]
table [row sep=\\]{%
0 0 \\
0.248764983698804	0.263169970631424 \\
0.497529967232926	0.35744768722467 \\
0.746294950767049	0.419557635829662 \\
0.995059934301171	0.476573972099853 \\
1.24382491783529	0.542469713656388 \\
1.49258990136942	0.630116556534508 \\
1.74135488490354	0.713839941262849 \\
1.99011986843766	0.766221549192364 \\
2.23888485197178	0.823788546255507 \\
2.4876498355059	0.843795888399413 \\
2.73641481904003	0.856552863436123 \\
2.98517980257415	0.869539280469897 \\
3.23394478610827	0.880781938325991 \\
3.48270976964239	0.890602055800294 \\
3.73147475317652	0.899665014684288 \\
3.98023973671064	0.907580763582966 \\
4.22900472024476	0.915748898678414 \\
4.47776970377888	0.923113986784141 \\
4.726534687313	0.927909324522761 \\
4.97529967084713	0.93357654185022 \\
5.22406465438125	0.93880781938326 \\
5.47282963791537	0.943419603524229 \\
5.72159462144949	0.94771016886931 \\
5.97035960498362	0.952826725403818 \\
6.21912458851774	0.957966226138032 \\
6.46788957205186	0.962279735682819 \\
6.71665455558598	0.965216593245228 \\
6.9654195391201	0.967075073421439 \\
7.21418452265423	0.969300660792952 \\
7.46294950618835	0.970906754772394 \\
7.71171448972247	0.972168685756241 \\
7.96047947325659	0.97372889133627 \\
8.20924445679072	0.975334985315712 \\
8.45800944032484	0.977262298091043 \\
8.70677442385896	0.979029001468429 \\
8.95553940739308	0.980726872246696 \\
9.20430439092721	0.982287077826725 \\
9.45306937446133	0.984489720998532 \\
9.70183435799545	0.986623531571219 \\
9.95059934152957	0.988573788546256 \\
10.1993643250637	0.990340491923642 \\
10.4481293085978	0.991831864904552 \\
10.6968942921319	0.99327734948605 \\
10.9456592756661	0.994952276064611 \\
11.1944242592002	0.99607654185022 \\
11.4431892427343	0.99720080763583 \\
11.6919542262684	0.998370961820852 \\
11.9407192098026	0.99912812041116 \\
12.1894841933367	0.999678781204111 \\
12.4382491768708	1 \\
};




\end{semilogxaxis}

\end{tikzpicture}}%
	\caption{The distribution of the effective throughput for Scenario C under different band switch polices.} 
	\label{fig:results_sc_c}
\end{figure}
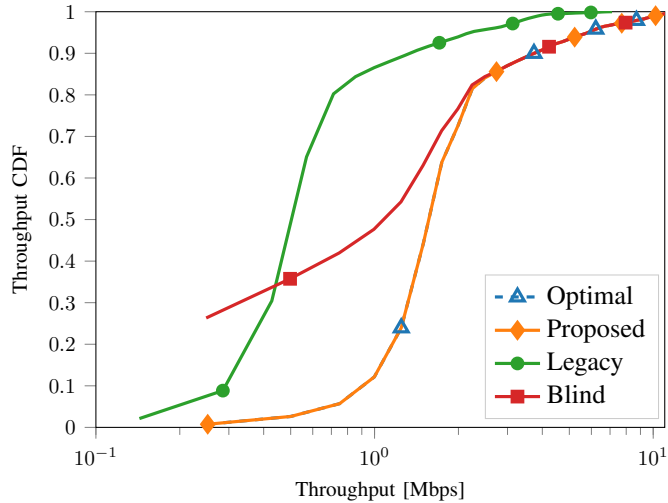

The results for this scenario are presented in Fig.~\ref{fig:results_sc_c}. As expected, the results \review{lie} between the previous two, and all the curves can be justified using the same arguments we \review{above}. The reason we include this scenario is to have an idea on the gains we might observe in practice, since part of the users will be using mmWave and the others using sub-6 GHz. The mean gains are also presented in Table~\ref{table:ho_req_granted}.

Overall, the results for the different scenarios show the superiority of the proposed policy compared to the legacy and the blind policies; up to $130\%$ improvement in the effective throughput depending on the considered scenario. It also justifies the use of a machine learning approach to solve this problem.  Next, we provide more technical discussions on the accuracy of the proposed algorithm, more insights, and possible extensions to this work.

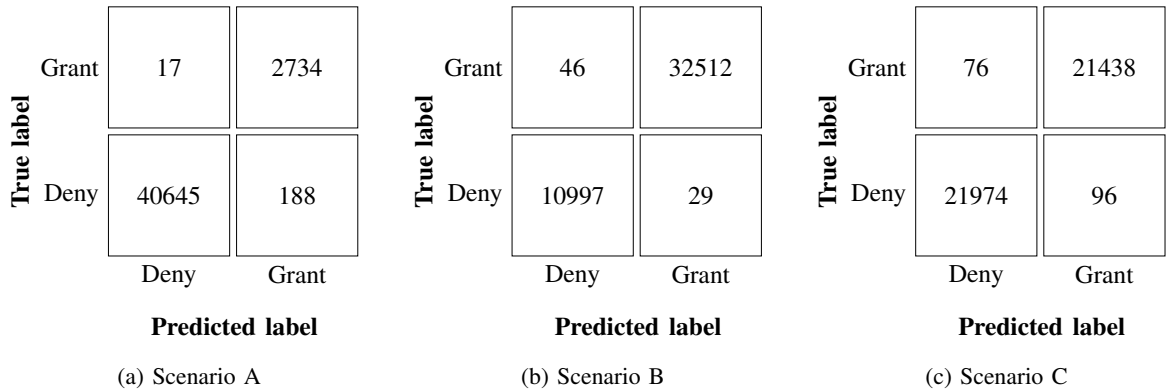
\begin{figure}[!t]
    \centering
    \subfloat[Scenario A]{\resizebox{0.3\textwidth}{!}{\begin{tikzpicture}[
box/.style={draw,rectangle,minimum size=2cm,text width=1.5cm,align=center}]
\matrix (conmat) [row sep=.1cm,column sep=.1cm] {
\node (tpos) [box,
    label=left:\(  \text{Grant} \),
    ] {17};
&
\node (fneg) [box
    ] {2734};
\\
\node (fpos) [box,
    label=left:\( \text{Deny} \),
    label=below:\(\text{Deny}\)] {40645};
&
\node (tneg) [box,
    label=below:\(\text{Grant}\)
    ] {188};
\\
};
\node [rotate=90, left=.05cm of conmat,anchor=center,text width=3cm,align=center] {\textbf{True label}};
\node [below=.05cm of conmat,xshift=.65cm,align=center] {\textbf{Predicted label}};
\end{tikzpicture}}}%
     \hfil
    \subfloat[Scenario B]{\resizebox{0.3\textwidth}{!}{\begin{tikzpicture}[
box/.style={draw,rectangle,minimum size=2cm,text width=1.5cm,align=center}]
\matrix (conmat) [row sep=.1cm,column sep=.1cm] {
\node (tpos) [box,
    label=left:\(  \text{Grant} \),
    ] {46};
&
\node (fneg) [box
    ] {32512};
\\
\node (fpos) [box,
    label=left:\( \text{Deny} \),
    label=below:\(\text{Deny}\)] {10997};
&
\node (tneg) [box,
    label=below:\(\text{Grant}\)
    ] {29};
\\
};
\node [rotate=90, left=.05cm of conmat,anchor=center,text width=3cm,align=center] {\textbf{True label}};
\node [below=.05cm of conmat,xshift=.65cm,align=center] {\textbf{Predicted label}};
\end{tikzpicture}}}%
    \hfil
    \subfloat[Scenario C]{\resizebox{0.3\textwidth}{!}{\begin{tikzpicture}[
box/.style={draw,rectangle,minimum size=2cm,text width=1.5cm,align=center}]
\matrix (conmat) [row sep=.1cm,column sep=.1cm] {
\node (tpos) [box,
    label=left:\(  \text{Grant} \),
    ] {76};
&
\node (fneg) [box
    ] {21438};
\\
\node (fpos) [box,
    label=left:\( \text{Deny} \),
    label=below:\(\text{Deny}\)] {21974};
&
\node (tneg) [box,
    label=below:\(\text{Grant}\)
    ] {96};
\\
};
\node [rotate=90, left=.05cm of conmat,anchor=center,text width=3cm,align=center] {\textbf{True label}};
\node [below=.05cm of conmat,xshift=.65cm,align=center] {\textbf{Predicted label}};
\end{tikzpicture}}}%
    \hfil
    \caption{The confusion matrix $\mathbf{C}$ for the three scenarios using DNN.} 
    \label{fig:conf_mat_dnn}
\end{figure}

\begin{figure}[!t]
    \centering
    \subfloat[Scenario A]{\resizebox{0.3\textwidth}{!}{\begin{tikzpicture}[
box/.style={draw,rectangle,minimum size=2cm,text width=1.5cm,align=center}]
\matrix (conmat) [row sep=.1cm,column sep=.1cm] {
\node (tpos) [box,
    label=left:\(  \text{Grant} \),
    ] {48};
&
\node (fneg) [box
    ] {2703};
\\
\node (fpos) [box,
    label=left:\( \text{Deny} \),
    label=below:\(\text{Deny}\)] {40651};
&
\node (tneg) [box,
    label=below:\(\text{Grant}\)
    ] {182};
\\
};
\node [rotate=90, left=.05cm of conmat,anchor=center,text width=3cm,align=center] {\textbf{True label}};
\node [below=.05cm of conmat,xshift=.65cm,align=center] {\textbf{Predicted label}};
\end{tikzpicture}}}%
     \hfil
    \subfloat[Scenario B]{\resizebox{0.3\textwidth}{!}{\begin{tikzpicture}[
box/.style={draw,rectangle,minimum size=2cm,text width=1.5cm,align=center}]
\matrix (conmat) [row sep=.1cm,column sep=.1cm] {
\node (tpos) [box,
    label=left:\(  \text{Grant} \),
    ] {208};
&
\node (fneg) [box
    ] {32350};
\\
\node (fpos) [box,
    label=left:\( \text{Deny} \),
    label=below:\(\text{Deny}\)] {10915};
&
\node (tneg) [box,
    label=below:\(\text{Grant}\)
    ] {111};
\\
};
\node [rotate=90, left=.05cm of conmat,anchor=center,text width=3cm,align=center] {\textbf{True label}};
\node [below=.05cm of conmat,xshift=.65cm,align=center] {\textbf{Predicted label}};
\end{tikzpicture}}}%
    \hfil
    \subfloat[Scenario C]{\resizebox{0.3\textwidth}{!}{\begin{tikzpicture}[
box/.style={draw,rectangle,minimum size=2cm,text width=1.5cm,align=center}]
\matrix (conmat) [row sep=.1cm,column sep=.1cm] {
\node (tpos) [box,
    label=left:\(  \text{Grant} \),
    ] {103};
&
\node (fneg) [box
    ] {21411};
\\
\node (fpos) [box,
    label=left:\( \text{Deny} \),
    label=below:\(\text{Deny}\)] {21907};
&
\node (tneg) [box,
    label=below:\(\text{Grant}\)
    ] {163};
\\
};
\node [rotate=90, left=.05cm of conmat,anchor=center,text width=3cm,align=center] {\textbf{True label}};
\node [below=.05cm of conmat,xshift=.65cm,align=center] {\textbf{Predicted label}};
\end{tikzpicture}}}%
    \hfil
    \caption{The confusion matrix $\mathbf{C}$ for the three scenarios using XGBoost.} 
    \label{fig:conf_mat_xgb}
\end{figure}
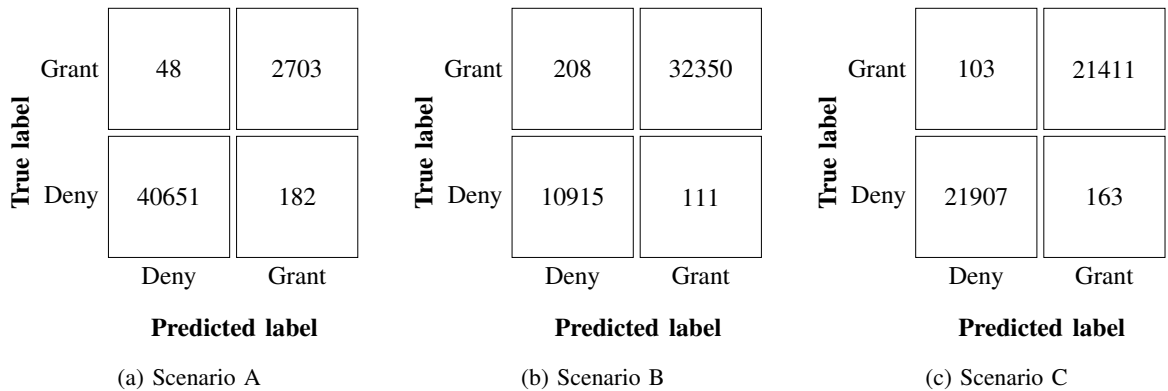

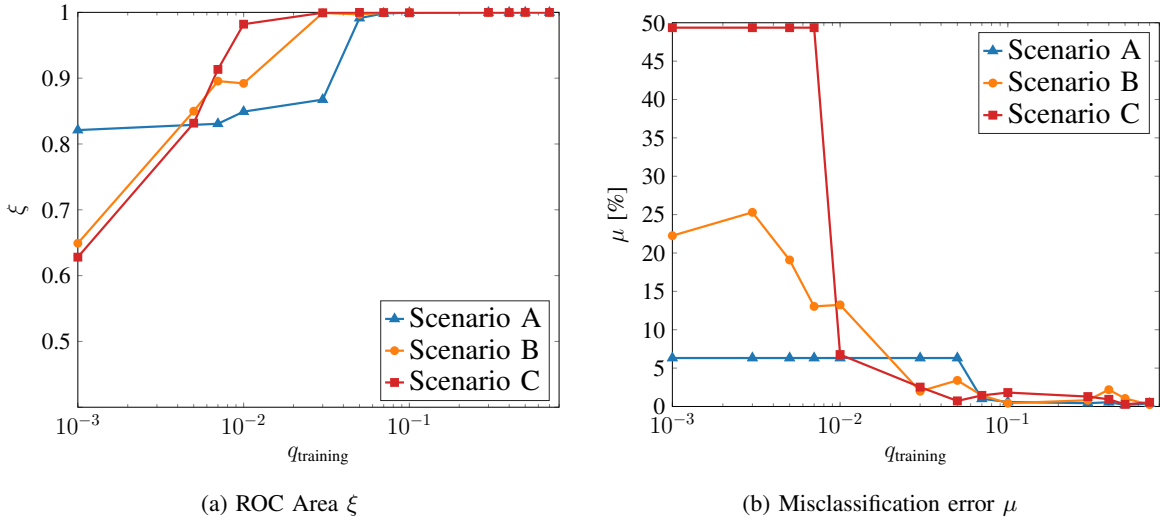
\begin{figure}[!t]
    \centering
    \subfloat[ROC Area $\xi$]{\resizebox{0.45\textwidth}{!}{\begin{tikzpicture}
\tikzstyle{every node}=[font=\large,scale=1.5]

\definecolor{color0}{rgb}{0.12156862745098,0.466666666666667,0.705882352941177}
\definecolor{color1}{rgb}{1,0.498039215686275,0.0549019607843137}
\definecolor{color3}{rgb}{0.83921568627451,0.152941176470588,0.156862745098039}

\begin{semilogxaxis}[%
width=6.028in,
height=4.95in,
at={(1.011in,0.642in)},
scale only axis,
xmin=0.001,
xmax=0.8,
grid style={white!69.01960784313725!black, dashed},
xlabel={$q_\text{training}$},
ymin=0.4,
ymax=1,
ytick={0.4,0.5,0.6,0.7,0.8,0.9,1},
yticklabels={,0.5,0.6,0.7,0.8,0.9,1},
ylabel={$\xi$},
axis background/.style={fill=white},
title style={font=\bfseries},
legend style={at={(0.98,0.27)},legend cell align=left,align=left,draw=white!15!black, nodes={scale=1.25, transform shape}},
]
\addplot [line width=2pt, color=color0,  mark size=4, mark=triangle*, mark options={solid}]
table [row sep=\\]{%
0.001	0.821100847479377 \\
0.007	0.830711857768443 \\
0.01	0.84913563000354 \\
0.03	0.86739967868164 \\
0.05	0.991195049748386 \\
0.07	0.998395526928522 \\
0.1	0.999528235972603 \\
0.3	0.999747132558436 \\
0.4	0.999892167459262 \\
0.5	0.999979329054768 \\
0.7	0.999837000427565 \\
};
\addlegendentry{Scenario A};

\addplot [line width=2pt, color=color1,  mark size=3, mark=*, mark options={solid}]
table [row sep=\\]{%
0.001	0.649119156417747 \\
0.005	0.849878869981765 \\
0.007	0.895561895668211 \\
0.01	0.892072488821718 \\
0.03	0.999121664046851 \\
0.05	0.996967448522876 \\
0.07	0.999302153729709 \\
0.1	0.99995617081058 \\
0.3	0.999774603086772 \\
0.4	0.999766775450934 \\
0.5	0.999809835804948 \\
0.7	0.999961795008714 \\
};
\addlegendentry{Scenario B};

\addplot [line width=2pt,color=color3, mark size=3, solid,mark=square*,mark options={solid}]
  table[row sep=crcr]{%
0.001	0.628053114611326 \\
0.005	0.831652360994089 \\
0.007	0.91332938743716 \\
0.01	0.982016705784442 \\
0.03	0.999359473366812 \\
0.05	0.999915282612361 \\
0.07	0.999724129858181 \\
0.1	0.999385171851932 \\
0.3	0.999944201305951 \\
0.4	0.999769118002802 \\
0.5	0.999983406133071 \\
0.7	0.999921864137193 \\
};
\addlegendentry{Scenario C};

\end{semilogxaxis}
\end{tikzpicture}%
}}%
    \hfil
    \subfloat[Misclassification error $\mu$]{\resizebox{0.45\textwidth}{!}{\begin{tikzpicture}
\tikzstyle{every node}=[font=\large,scale=1.5]

\definecolor{color0}{rgb}{0.12156862745098,0.466666666666667,0.705882352941177}
\definecolor{color1}{rgb}{1,0.498039215686275,0.0549019607843137}
\definecolor{color3}{rgb}{0.83921568627451,0.152941176470588,0.156862745098039}

\begin{semilogxaxis}[%
width=6.028in,
height=4.754in,
at={(1.011in,0.642in)},
scale only axis,
xmin=0.001,
xmax=0.8,
grid style={white!69.01960784313725!black, dashed},
xlabel={$q_\text{training}$},
ymin=0,
ymax=50,
ytick={0,5,10,...,50},
yticklabels={0,5,10,...,50},
ylabel={$\mu$ [\%]},
axis background/.style={fill=white},
title style={font=\bfseries},
legend style={at={(0.98,0.975)},legend cell align=left,align=left,draw=white!15!black, nodes={scale=1.25, transform shape}}
]       
\addplot [line width=2pt, color=color0,  mark size=4, mark=triangle*, mark options={solid}]
table [row sep=\\]{%
0.001	6.31194933920705 \\
0.003	6.31194933920705 \\
0.005	6.31194933920705 \\
0.007	6.31194933920705 \\
0.01	6.31194933920705 \\
0.03	6.31194933920705 \\
0.05	6.31194933920705 \\
0.07	0.995778267254038 \\
0.1	0.587371512481645 \\
0.3	0.429056534508076 \\
0.4	0.54836637298091 \\
0.5	0.204203377386197 \\
0.7	0.367107195301028 \\
};
\addlegendentry{Scenario A};

\addplot [line width=2pt, color=color1,  mark size=3, mark=*, mark options={solid}]
table [row sep=\\]{%
0.001	22.2535792951542 \\
0.003	25.2982745961821 \\
0.005	19.0895741556535 \\
0.007	13.0391886930984 \\
0.01	13.241097650514 \\
0.03	1.98237885462555 \\
0.05	3.39803597650514 \\
0.07	1.44089574155653 \\
0.1	0.424467694566813 \\
0.3	0.812224669603524 \\
0.4	2.18658223201175 \\
0.5	1.0439610866373 \\
0.7	0.2409140969163 \\
};
\addlegendentry{Scenario B};

\addplot [line width=2pt,color=color3, mark size=3, solid,mark=square*,mark options={solid}]
  table[row sep=crcr]{%
0.001	49.3621512481645 \\
0.003	49.3621512481645 \\
0.005	49.3621512481645 \\
0.007	49.3621512481645 \\
0.01	6.76165565345081 \\
0.03	2.52156754772394 \\
0.05	0.72503671071953 \\
0.07	1.44089574155653 \\
0.1	1.81259177679883 \\
0.3	1.28946402349486 \\
0.4	0.929240088105727 \\
0.5	0.277624816446402 \\
0.7	0.539188693098385 \\
};
\addlegendentry{Scenario C};

\end{semilogxaxis}
\end{tikzpicture}%
}}%
    \caption{The classification performance of the proposed algorithm for different training data sizes and different scenarios.}
    \label{fig:roc_misclass}
\end{figure}

\subsection{Discussion}

We start with discussing the predictive accuracy of the proposed algorithm. Fig.~\ref{fig:conf_mat_dnn} shows that our proposed algorithm \review{usually} made the right decisions; \review{only a very few times did it deny} the band switch when it was supposed to grant it (and vice versa).

To compare the performance of the ML classifiers\footnote{These comparisons may not be general for any ML algorithm, but are valid for these important ML classification algorithms.}, we show the performance of XGBoost alongside DNN. In Fig.~\ref{fig:conf_mat_dnn} and Fig.~\ref{fig:conf_mat_xgb}, we show the confusion matrix for the three considered scenarios using DNN and XGBoost respectively.  Precisely, the misclassification error ($\mu$) using DNN (XGBoost) is 0.47\% (0.53\%), 0.17\% (0.73\%), and 0.39\% (0.61\%) for Scenarios A, B, and C, respectively. The run-time complexity of XGBoost using the hyperparameters in Table~\ref{table:hyperparams_isho} has an upper bound in $\mathcal{O}(n + n\log n) = \mathcal{O}(n\log n)$ \cite{XGBoost_Chen16}, where $n := N_\text{learning}$.   However, for DNN this complexity is {super-linear since training DNNs requires matrix multiplications \cite{Deep_Goodfellow16}.  Matrix multiplications have a run-time complexity between $\mathcal{O}(n^2)$ and $\mathcal{O}(n^3)$}.  Hence, the classifier choice between DNN and XGBoost is a trade-off between decision speed and accuracy: if accuracy is desired, then choose DNN, but if less run-time complexity is desired for decision speed, XGBoost is a more attractive choice.

The second point we highlight here is the amount of training data that we require to have an accurate prediction. In Fig.~\ref{fig:roc_misclass}, we show the ROC area ($\xi$) and the misclassification error ($\mu$) for different training data sizes. In particular, the figure shows that training using only {$1$,$362$ measurements (i.e., $1/40$ of the data {for a grid of an area of 19,250 square meters}) is enough to have an excellent performance---less than $2\%$ misclassification error. In other words, having knowledge about the previous band switch decisions of {7 random samples per 100 square meters} is enough to predict the band switch decisions for the rest of the locations. This {absolute number} depends on the spatial correlation between the channels on different locations, as well as the hyperparameters and the choice of the machine learning algorithm.   {Further, this insight should be understood alongside the other considerations, such as the user grid size, the collection period, and the blockage probabilities, as discussed in this section.}

Note that our presented results so far are for a single band switch threshold value.  However, we claim that the performance gap between the optimal algorithm and the proposed one can be reduced by increasing the threshold. To verify  this  claim,  we  show  the  mean  effective  throughput  for  different  band switch  thresholds  in Table~\ref{table:scA_thresholds}.  In Scenario A, we observe that as we increase the band switch threshold $r_\text{threshold}$, the performance gap between the mean effective throughputs of the proposed and the optimal rates shrink considerably.  While both the legacy and blind rates also get better, their performance is not close to the optimal: the legacy because of the measurement gap and the blind because of the undesired band switch. However, in Scenario B, both the blind and legacy rates deteriorate as we increase the band switch thresholds.  In the legacy policy, it is also due to the measurement gaps, and for the blind policy, it is because users who were getting up to 10 Mbps on mmWave are now getting $3$ Mbps at best as shown in Fig.~\ref{fig:pdf}.  As expected, we do not see much of a change in the proposed rate as we increase the band switch threshold, aligned with the CDFs in Fig.~\ref{fig:Diff} and Fig.~\ref{fig:Sc_b}. 

{Further, to show the behavior of the proposed policy against uncertainty due to the change of the blockage probability as in \eqref{eq:blockage}, we simulate Scenario C using $p \in \{0.2, 0.4, 0.6, 0.8\}$ in the exploitation data, while the learning data is fixed at $p = 0.4$.  Then, we compute the mean effective throughput as shown in Fig.~\ref{fig:blockage}.  The classifier remains resilient against uncertainty of band-selective blockage contrary to the other policies (with a significance up to the third decimal).  This is due to the ability of the classifier to learn from the spatial relationship of the channels even with blockage as a result of: 1) coordinates being part of the learning features and 2) the relaxation of $r_\text{threshold}$ value in the learning phase.}

\begin{figure}[!t]
	\centering
	\scalebox{1}{
\begin{tikzpicture}
\tikzstyle{every node}=[font=\smaller,scale=0.8]
\definecolor{color0}{rgb}{0.12156862745098,0.466666666666667,0.705882352941177}
\definecolor{color1}{rgb}{1,0.498039215686275,0.0549019607843137}
\definecolor{color2}{rgb}{0.172549019607843,0.627450980392157,0.172549019607843}
\definecolor{color3}{rgb}{0.83921568627451,0.152941176470588,0.156862745098039}

\begin{semilogxaxis}[
width=3.6in,
height=2.8in,
legend cell align={left},
legend entries={{Optimal},{Proposed},{Legacy},{Blind}},
legend style={at={(0.99,0.03)}, anchor=south east, draw=white!80.0!black, nodes={scale=1.25, transform shape}},
tick align=outside,
tick pos=left,
x grid style={white!69.01960784313725!black,dashed},
xlabel={Throughput [Mbps]},
xmin=0.1, xmax=12,
y grid style={white!69.01960784313725!black,dashed},
ylabel={Throughput CDF},
ymin=0, ymax=1, 
ytick={0,0.1,0.2,...,1}
]
\addplot [color0, line width=1.2pt, dashed, mark=triangle, mark size=3,mark repeat=10, mark phase=5, mark options={solid}]
table [row sep=\\]{%
0.251923681032548	0.00426762114537445 \\
0.500624201355778	0.0158544419970631 \\
0.749324721679007	0.0371696035242291 \\
0.998025242002237	0.0762894640234949 \\
1.24672576232547	0.16322503671072 \\
1.4954262826487	0.305570851688693 \\
1.74412680297193	0.521888766519824 \\
1.99282732329515	0.676486784140969 \\
2.24152784361838	0.767529368575624 \\
2.49022836394161	0.840147760646109 \\
2.73892888426484	0.855841593245228 \\
2.98762940458807	0.869745778267254 \\
3.2363299249113	0.880827826725404 \\
3.48503044523453	0.890693832599119 \\
3.73373096555776	0.8997109030837 \\
3.98243148588099	0.907603707782673 \\
4.23113200620422	0.915794787077827 \\
4.47983252652745	0.923159875183554 \\
4.72853304685068	0.927909324522761 \\
4.97723356717391	0.933668318649046 \\
5.22593408749714	0.93880781938326 \\
5.47463460782037	0.943419603524229 \\
5.7233351281436	0.947733113069016 \\
5.97203564846683	0.952849669603524 \\
6.22073616879006	0.957966226138032 \\
6.46943668911329	0.962302679882526 \\
6.71813720943652	0.965239537444934 \\
6.96683772975975	0.967098017621145 \\
7.21553825008298	0.969323604992658 \\
7.46423877040621	0.9709296989721 \\
7.71293929072943	0.972168685756241 \\
7.96163981105266	0.97372889133627 \\
8.21034033137589	0.975357929515418 \\
8.45904085169912	0.977262298091042 \\
8.70774137202235	0.979029001468429 \\
8.95644189234558	0.980726872246696 \\
9.20514241266881	0.982287077826725 \\
9.45384293299204	0.984489720998531 \\
9.70254345331527	0.986623531571219 \\
9.9512439736385	0.988573788546255 \\
10.1999444939617	0.990340491923641 \\
10.448645014285	0.991831864904552 \\
10.6973455346082	0.99327734948605 \\
10.9460460549314	0.994952276064611 \\
11.1947465752546	0.99607654185022 \\
11.4434470955779	0.99720080763583 \\
11.6921476159011	0.998370961820852 \\
11.9408481362243	0.99912812041116 \\
12.1895486565476	0.999678781204112 \\
12.4382491768708	1 \\
};
\addplot [line width=1.2pt, color1, mark=diamond*, mark size=3, mark phase=1, mark repeat=10, mark options={solid}]
table [row sep=\\]{%
0.251923681032548	0.00426762114537445 \\
0.500624201355778	0.0158544419970631 \\
0.749324721679007	0.0371696035242291 \\
0.998025242002237	0.0762894640234949 \\
1.24672576232547	0.16322503671072 \\
1.4954262826487	0.305639684287812 \\
1.74412680297193	0.521980543318649 \\
1.99282732329515	0.676739170337739 \\
2.24152784361838	0.767758810572687 \\
2.49022836394161	0.840147760646109 \\
2.73892888426484	0.855841593245228 \\
2.98762940458807	0.869745778267254 \\
3.2363299249113	0.880827826725404 \\
3.48503044523453	0.890693832599119 \\
3.73373096555776	0.8997109030837 \\
3.98243148588099	0.907603707782673 \\
4.23113200620422	0.915794787077827 \\
4.47983252652745	0.923159875183554 \\
4.72853304685068	0.927909324522761 \\
4.97723356717391	0.933668318649046 \\
5.22593408749714	0.93880781938326 \\
5.47463460782037	0.943419603524229 \\
5.7233351281436	0.947733113069016 \\
5.97203564846683	0.952849669603524 \\
6.22073616879006	0.957966226138032 \\
6.46943668911329	0.962302679882526 \\
6.71813720943652	0.965239537444934 \\
6.96683772975975	0.967098017621145 \\
7.21553825008298	0.969323604992658 \\
7.46423877040621	0.9709296989721 \\
7.71293929072943	0.972168685756241 \\
7.96163981105266	0.97372889133627 \\
8.21034033137589	0.975357929515418 \\
8.45904085169912	0.977262298091043 \\
8.70774137202235	0.979029001468429 \\
8.95644189234558	0.980726872246696 \\
9.20514241266881	0.982287077826725 \\
9.45384293299204	0.984489720998531 \\
9.70254345331527	0.986623531571219 \\
9.9512439736385	0.988573788546255 \\
10.1999444939617	0.990340491923642 \\
10.448645014285	0.991831864904552 \\
10.6973455346082	0.99327734948605 \\
10.9460460549314	0.994952276064611 \\
11.1947465752546	0.99607654185022 \\
11.4434470955779	0.99720080763583 \\
11.6921476159011	0.998370961820852 \\
11.9408481362243	0.99912812041116 \\
12.1895486565476	0.999678781204112 \\
12.4382491768708	1 \\
};
\addplot [line width=1.2pt, color2, mark=*, mark size=2, mark phase=3, mark repeat=5, mark options={solid}]
table [row sep=\\]{%
0.249736244856994	0.048481093979442 \\
0.498481406734826	0.464757709251101 \\
0.747226568612659	0.836683186490455 \\
0.995971730490492	0.880988436123348 \\
1.24471689236832	0.91017345814978 \\
1.49346205424616	0.930892070484581 \\
1.74220721612399	0.946746512481645 \\
1.99095237800182	0.962348568281938 \\
2.23969753987965	0.962623898678414 \\
2.48844270175749	0.962623898678414 \\
2.73718786363532	0.962623898678414 \\
2.98593302551315	0.962623898678414 \\
3.23467818739099	0.962623898678414 \\
3.48342334926882	0.962623898678414 \\
3.73216851114665	0.962623898678414 \\
3.98091367302448	0.962623898678414 \\
4.22965883490232	0.962623898678414 \\
4.47840399678015	0.962623898678414 \\
4.72714915865798	0.962623898678414 \\
4.97589432053582	0.962623898678414 \\
5.22463948241365	0.962623898678414 \\
5.47338464429148	0.962623898678414 \\
5.72212980616931	0.962623898678414 \\
5.97087496804715	0.962623898678414 \\
6.21962012992498	0.962623898678414 \\
6.46836529180281	0.962623898678414 \\
6.71711045368064	0.965216593245228 \\
6.96585561555848	0.967075073421439 \\
7.21460077743631	0.969300660792952 \\
7.46334593931414	0.970906754772394 \\
7.71209110119197	0.972168685756241 \\
7.96083626306981	0.97372889133627 \\
8.20958142494764	0.975357929515419 \\
8.45832658682547	0.977262298091043 \\
8.7070717487033	0.979029001468429 \\
8.95581691058114	0.980726872246696 \\
9.20456207245897	0.982287077826725 \\
9.4533072343368	0.984489720998531 \\
9.70205239621463	0.986623531571219 \\
9.95079755809247	0.988573788546255 \\
10.1995427199703	0.990340491923642 \\
10.4482878818481	0.991831864904552 \\
10.697033043726	0.99327734948605 \\
10.9457782056038	0.994952276064611 \\
11.1945233674816	0.99607654185022 \\
11.4432685293595	0.99720080763583 \\
11.6920136912373	0.998370961820852 \\
11.9407588531151	0.99912812041116 \\
12.189504014993	0.999678781204112 \\
12.4382491768708	1 \\
};
\addplot [line width=1.2pt, color3, mark=square*, mark size=2, mark phase=1, mark repeat=15, mark options={solid}]
table [row sep=\\]{%
0.250614762259805	0.00718153450807636 \\
0.49934199521105	0.0196861233480176 \\
0.748069228162294	0.0440758076358297 \\
0.996796461113539	0.0874403450807636 \\
1.24552369406478	0.177565161527166 \\
1.49425092701603	0.322228340675477 \\
1.74297815996727	0.549444750367107 \\
1.99170539291852	0.722512848751836 \\
2.24043262586976	0.849508994126285 \\
2.48915985882101	0.961499632892805 \\
2.73788709177225	0.962623898678414 \\
2.9866143247235	0.962623898678414 \\
3.23534155767474	0.962623898678414 \\
3.48406879062599	0.962623898678414 \\
3.73279602357723	0.962623898678414 \\
3.98152325652848	0.962623898678414 \\
4.23025048947972	0.962623898678414 \\
4.47897772243097	0.962623898678414 \\
4.72770495538221	0.962623898678414 \\
4.97643218833345	0.962623898678414 \\
5.2251594212847	0.962623898678414 \\
5.47388665423594	0.962623898678414 \\
5.72261388718719	0.962623898678414 \\
5.97134112013843	0.962623898678414 \\
6.22006835308968	0.962623898678414 \\
6.46879558604092	0.962623898678414 \\
6.71752281899217	0.965216593245228 \\
6.96625005194341	0.967075073421439 \\
7.21497728489466	0.969300660792952 \\
7.4637045178459	0.970906754772394 \\
7.71243175079715	0.972168685756241 \\
7.96115898374839	0.97372889133627 \\
8.20988621669964	0.975357929515419 \\
8.45861344965088	0.977262298091043 \\
8.70734068260212	0.979029001468429 \\
8.95606791555337	0.980726872246696 \\
9.20479514850461	0.982287077826725 \\
9.45352238145586	0.984489720998532 \\
9.7022496144071	0.986623531571219 \\
9.95097684735835	0.988573788546255 \\
10.1997040803096	0.990340491923642 \\
10.4484313132608	0.991831864904552 \\
10.6971585462121	0.99327734948605 \\
10.9458857791633	0.994952276064611 \\
11.1946130121146	0.99607654185022 \\
11.4433402450658	0.997200807635829 \\
11.6920674780171	0.998370961820851 \\
11.9407947109683	0.99912812041116 \\
12.1895219439195	0.999678781204111 \\
12.4382491768708	1 \\
};




\end{semilogxaxis}

\end{tikzpicture}}%
	\caption{The distribution of the effective throughput for Scenario B under different band switch polices.}
	\label{fig:Sc_b}
\end{figure}
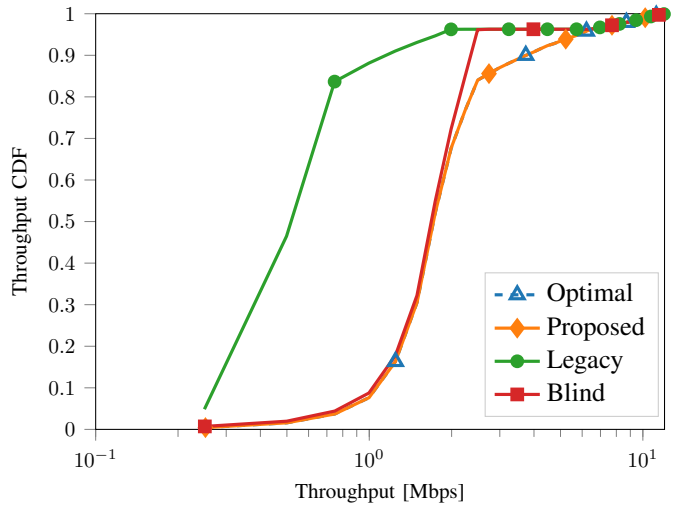 


\begin{figure}[!t]
	\centering
	{
    	\scalebox{1}{\begin{tikzpicture}
\tikzstyle{every node}=[font=\smaller,scale=0.8]

\definecolor{color0}{rgb}{0.12156862745098,0.466666666666667,0.705882352941177}
\definecolor{color1}{rgb}{1,0.498039215686275,0.0549019607843137}
\definecolor{color2}{rgb}{0.172549019607843,0.627450980392157,0.172549019607843}
\definecolor{color3}{rgb}{0.83921568627451,0.152941176470588,0.156862745098039}

\begin{axis}[
width=3.6in,
height=2.8in,
legend cell align={left},
legend entries={{Proposed},{Legacy},{Blind}},
legend style={at={(0.02,0.23)}, anchor=south west, draw=white!80.0!black, nodes={scale=1.25, transform shape}},
tick align=outside,
tick pos=left,
x grid style={white!69.01960784313725!black,dashed},
xlabel={$p$},
xmin=0.1, xmax=0.9,
y grid style={white!69.01960784313725!black,dashed},
ylabel={$R_E$ (Normalized to optimal)},
ymin=0.3, ymax=1.1, 
ytick={0.2,0.4,...,1.0}
]

\addplot [line width=1.2pt, color1, mark=diamond*, mark size=3, mark phase=1, mark repeat=1, mark options={solid}]
table [row sep=\\]{%
0.2 0.999915454\\
0.4 0.999910407 \\
0.6 0.999898496\\
0.8 0.999891356 \\
};
\addplot [line width=1.2pt, color2, mark=*, mark size=2, mark phase=1, mark repeat=1, mark options={solid}]
table [row sep=\\]{%
0.2 0.415650301	\\
0.4 0.402425003	\\
0.6 0.387289657 \\
0.8 0.370600704 \\
};
\addplot [line width=1.2pt, color3, mark=square*, mark size=2, mark phase=1, mark repeat=1, mark options={solid}]
table [row sep=\\]{%
 0.2 0.754312832 \\
 0.4 0.712447713 \\
 0.6 0.660170809 \\
 0.8 0.600245289 \\
};




\end{axis}
\end{tikzpicture}}%
    	\caption{Normalized mean effective throughput for different blockage probabilities $p$ in the \textit{exploitation} phase }
    	\label{fig:blockage}
	}
\end{figure}
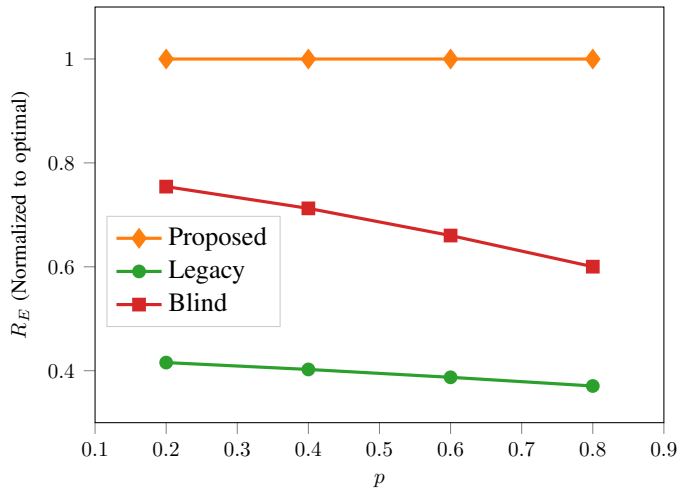 

\begin{table}[!t]
    \setlength\doublerulesep{0.5pt}
    \caption{Impact of different band switch thresholds for Scenarios A and B}
    \label{table:scA_thresholds}
    \vspace*{-1em}
    \centering
        \begin{tabular}{lc|c@{\hskip 0.5in}c@{\hskip 0.5in}c@{\hskip 0.5in}c}
        \hhline{======}
         & & \multicolumn{4}{c}{\review{Normalized} mean effective throughput $R_E$} \\
         \cline{3-6}
        & $r_\text{threshold}$ & Legacy & Blind & Proposed & Optimal \\
        \hline
        
& 1.72 & 0.55 & 0.54 & 0.75  & 1.00 \\
Scenario A & 2.00 & 0.45 & 0.46 & 0.77 & 1.00 \\ 
& 2.60 & 0.34 & 0.60 & 1.00 & 1.00 \\

        \hline
& 2.00 & 0.43 & 0.88 & 1.00 & 1.00 \\
Scenario B & 9.00 & 0.39 & 0.84 & 1.00 & 1.00 \\
& 12.50 & 0.33 & 0.76 & 1.00 & 1.00 \\
        \hhline{======}
        \end{tabular}%
\end{table}

\section{Conclusions}\label{sec:conclusion}
In this paper, we used both deep neural networks and XGBoost classifiers to rank the downlink channel quality of the frequency bands prior to the band switch, which is a mathematically intractable problem. The use of classifiers {in an online learning setting} eliminates the dependence on measurement gaps during a band switch in a dual-band base station.  We exploited the spatial and spectral relationships in both the sub-6 GHz and mmWave bands through the use of a ray-tracing dataset. {This brought forward two benefits: 1) reduces link latency by removing the need for a measurement gap and 2) reduces complexity in the UE and BS because channel estimation in the other frequency band is not required.} We revealed insights as to why the deep learning classification method was needed and why it worked.  We simulated one dual-band base station with many UEs in its association area {and varied the blockage probability}.  In this simulation, our method improved downlink throughput by up to $1.3\text{x}$ compared with the legacy policy over different scenarios with a misclassification error less than 0.3\%. The observed improvement is due to the classifier ability to exploit the spatial correlation of channels across the different frequency bands and thus accurately predict the effective achievable rate on the target frequency without the dependency on a measurement gap.  This band selection method is better suited for 5G and beyond where maintaining high data rates is desired without interrupting the data flow.  We focused on the case where the BS has only two bands: one centered at 3.5 GHz and the other at 28 GHz, since the dataset we use supports these two bands. An interesting \review{extension is for} multiple bands, or when a handoff between multiple BSs is required  {due to mobility, when subsequent band switches are required (also known as the ``ping-pong'' effect)}, {or when the BS uses different radio units per frequency band as in \cite{8642794}.} 

\bibliographystyle{IEEEtran}
\bibliography{RefUpdated.bib}  
\end{document}